\documentclass[aps,reprint,amsmath,amssymb,twocolumnm]{revtex4-2}
\usepackage{aas_macros}
\usepackage{natbib}
\usepackage{color}
\usepackage{graphicx}
\usepackage{dcolumn}
\usepackage{bm}

\begin{document}

\title{A Monte-Carlo based relativistic radiation hydrodynamics code with a higher-order scheme}

\author{Kyohei Kawaguchi}
\affiliation{Institute for Cosmic Ray Research, The University of Tokyo, 5-1-5 Kashiwanoha, Kashiwa, Chiba 277-8582, Japan}
\affiliation{Center of Gravitational Physics and Quantum Information,
 Yukawa Institute for Theoretical Physics, 
Kyoto University, Kyoto, 606-8502, Japan} 
\author{Sho Fujibayashi}
\affiliation{Max Planck Institute for Gravitational Physics (Albert Einstein Institute), Am M\"{u}hlenberg 1, Potsdam-Golm, 14476, Germany}
\author{Masaru Shibata}
\affiliation{Max Planck Institute for Gravitational Physics (Albert Einstein Institute), Am M\"{u}hlenberg 1, Potsdam-Golm, 14476, Germany}
\affiliation{Center of Gravitational Physics and Quantum Information,
 Yukawa Institute for Theoretical Physics, 
Kyoto University, Kyoto, 606-8502, Japan} 

\newcommand{\angstrom}{\text{\normalfont\AA}}
\newcommand{\rednote}[1]{{\color{red} (#1)}}
\newcommand{\ms}[1]{{\color{blue} #1}}
\newcommand{\kk}[1]{{\color{magenta} #1}}

\begin{abstract}
We develop a new relativistic radiation hydrodynamics code based on the Monte-Carlo algorithm. In this code, we implement a new scheme to achieve the second-order accuracy in time in the limit of a large packet number for solving the interaction between matter and radiation. This higher-order time integration scheme is implemented in the manner to guarantee the energy-momentum conservation to the precision of the geodesic integrator. The spatial dependence of radiative processes, such as the packet propagation, emission, absorption, and scattering, are also taken into account up to the second-order accuracy. We validate our code by solving various test-problems following the previous studies; one-zone thermalization, dynamical diffusion, radiation dragging, radiation mediated shock-tube, shock-tube in the optically thick limit, and Eddington limit problems. We show that our code reproduces physically appropriate results with reasonable accuracy and also demonstrate that the second-order accuracy in time and space is indeed achieved with our implementation for one-zone and one-dimensional problems. 
\end{abstract}


\maketitle

\section{Introduction}\label{sec:intro}
The merger of neutron stars is one of the most interesting multi-messenger phenomena in high-energy astrophysics, in which physical processes in extreme (strongly self-gravitating, high-density, and high-temperature) environments are realized. The simultaneous detection of gravitational waves (GWs) from a binary neutron star and its electromagnetic (EM) counterparts provides a great opportunity to study such systems. Indeed, the first detection of GWs and EM signals from a binary neutron star, GW170817~\citep{2017PhRvL.119p1101A,LIGOScientific:2017ync},  demonstrates the powerfulness of the multi-messenger astronomy. A number of detections of GWs and EM signals from binary neutron stars are expected in the next few years~\citep{2018LRR....21....3A,2020MNRAS.493.1633S,2020arXiv200802921K,2022ApJS..260...18A}, and the observation of GWs and EM signals from a binary neutron star will surely give a great impact on both astrophysics and fundamental physics.

To maximize the scientific returns from the observed signals, the quantitative prediction of the merger outcome is crucial. After the binary merger, a massive neutron star or a black hole surrounded by a strongly magnetized hot and dense accretion torus is likely to be formed~\citep{2006Sci...312..719P,2018PhRvD..97l4039K}. The accretion torus is considered to launch a relativistic jet and outflows by magnetic pressure and tension, viscous heating due to magneto-hydrodynamical turbulence, and neutrino irradiation (e.g., Refs.~\cite{1977MNRAS.179..433B,BP82,Balbus:1998ja,Dessart:2008zd,2013MNRAS.435..502F}). In such a situation, neutrino-antineutrino pair annihilation could be the important mechanism for the system to launch a jet powerful enough to explain gamma-ray bursts~\citep{1996A&A...305..839J,1999ApJ...518..356P}. In addition to the matter ejected by tidal disruption and collisional shock heating at the onset of the merger (e.g., Refs.~\cite{Rosswog:1998hy,Ruffert:2001gf,Hotokezaka:2012ze}), the neutron-rich matter ejected in the post-merger phase is expected to be the important site of the $r$-process nucleosynthesis in the universe~\citep{1974ApJ...192L.145L,1989Natur.340..126E,1999ApJ...525L.121F,2021RvMP...93a5002C}. Since weak interaction processes play an important role in determining the dynamics and the thermodynamic properties of the merger remnants, the post-merger environment, and the abundance of the elements synthesized in the ejecta (e.g., Refs.~\cite{Metzger:2010sy,Goriely:2010bm,Wanajo:2014wha,Just:2014fka,Sekiguchi:2015dma,Sekiguchi:2016bjd,Radice:2016dwd,2019PhRvD.100b3008M,Fujibayashi:2017puw,Fujibayashi:2020qda,Fujibayashi:2020jfr,Fujibayashi:2020dvr,Foucart:2020qjb,Just:2021cls,2021arXiv211104621H,2022arXiv220505557F,2022arXiv220710680F}), accurately solving neutrino radiation is a key ingredient for the quantitative understanding of the merger physics.

While radiation of neutrinos and photons plays an important role in various astrophysical situation, solving radiative transfer is for many cases computationally expensive due to its large dimensionality of the phase space dependence; seven dimensions which come from time, 3 real-space dimensions, and 3 momentum-space dimensions. Moreover, the physical time scale of the local radiation-matter coupling can be often much shorter than the dynamical time scale of the matter field, and hence, complicated prescriptions, such as implicit solvers, are required to numerically solve the system in realistic computational time. The recent dramatic progress of computer resources has made it possible to directly solve radiation-transfer equations by the full discretization of a radiation field (e.g., Refs.~\cite{2014ApJS..214...16N,2017ApJS..229...42N,2014ApJS..213....7J,2022arXiv220906240J,2016ApJ...818..162O,2020ApJ...901...96A}), but yet, the size and resolution of the problems that can be solved are still limited. 

Various approximation methods are proposed for overcoming such problems. One of the most successful approximation methods among them is the moment scheme. In a moment scheme, up to the two lowest moments of radiation in momentum space are solved as the dynamical variables with an approximate closure relation to the higher moments~\citep{1980RvMP...52..299T,2011PThPh.125.1255S}. In the context of relativistic problems, many numerical codes are developed by employing moment schemes sometimes with a combination of the leakage algorithm~\citep{2013ApJ...772..127T,2014MNRAS.439..503S,2014MNRAS.441.3177M,2015MNRAS.447...49S,Sekiguchi:2015dma,2015PhRvD..91l4021F,Sekiguchi:2016bjd,Radice:2016dwd,2022arXiv220504487K}. However, while the moment schemes should be accurate for the optically thick cases, it sometimes fails to capture a physically correct property in the mildly optically thick or optically thin regions (e.g., see the result of the two-beam crossing problem presented in Ref.~\cite{2020ApJ...901...96A}).
Since the moment schemes do not necessarily provide a solution which converges to the correct solution of the full radiation-transfer equations, it is not guaranteed that the outcome derived from the moment schemes is always reliable. In fact, Ref.~\cite{2018PhRvD..98f3007F} points out that a moment scheme can underestimate the neutrino-antineutrino pair annihilation rate in a neutron star merger simulation.
 
An alternative approach for solving the radiation-transfer equation is the Monte-Carlo radiation transport. In the Monte-Carlo scheme, a radiation field is described by a set of packets, each of which represents a large number of photons or neutrinos, and its evolution is determined by solving the transport along the geodesic and by taking the interaction with the matter field into account for each packet. In the limit of the large packet number (ultimately which resolves each photon or neutrino), the solution obtained by the Monte-Carlo scheme manifestly converges to the solution of radiation-transfer equation. Recently, radiation hydrodynamics codes based on the Monte-Carlo scheme are developed by several groups~\citep{2012ApJ...755..111A,2015ApJS..217....9R,2015ApJ...807...31R,2018MNRAS.475.4186F,2019PhRvD.100b3008M,2019ApJS..241...30M,2021ApJ...920...82F,2022ApJ...933..226R} because the frequency dependence and complicated angular dependence expected in an optically thin region, as well as relativistic effects, can be incorporated in a straightforward manner. However, there are several drawbacks to the Monte-Carlo approach. Among them, the slow convergence of the statistical error of the Monte-Carlo packets (``the Monte-Carlo shot noise'') is a problem to be improved for the radiation-hydrodynamics codes with the Monte-Carlo algorithm. The operator splitting method is often used for the interaction between matter and radiation fields in all the previous studies, by which the error due to the finite discretization shows only the first order convergence.

In this paper, we report our new  Monte-Carlo-based radiation-hydrodynamics code as an improved version of the previous codes~\citep{2009ApJS..184..387D,2012ApJ...755..111A,2015ApJS..217....9R,2015ApJ...807...31R,2018MNRAS.475.4186F}. In this code, we implement a new scheme to achieve the second-order accuracy in time in the limit of a large packet number for solving the interaction between matter and radiation fields. We also take the spatial dependence of radiative processes into account up to the second-order accuracy, by which our code is second-order accurate in both time and space. We also propose a prescription which can be used for a very optically thick regime to suppress the Monte-Carlo shot noise of the energy-momentum transport between the cells. Our code is primarily designed to solve an axisymmetric system aiming at the long-term evolution of the post-merger phase (e.g., Ref.~\cite{Fujibayashi:2020dvr}), although the modification to a solution of the fully three-dimensional problems is straightforward. 

This paper is organized as follows: In Sec.~\ref{sec:basic}, we describe the basic formulation of the radiation hydrodynamics. In Sec.~\ref{sec:method}, we describe the numerical method employed in our code. In Sec.~\ref{sec:test}, we present successful results for various numerical test problems and validate our new code. Finally, we summarize our present work in Sec.~\ref{sec:sum}. Throughout this paper, $c$ and $G$ denote the speed of light and gravitational constant, respectively, and the units of $c=G=1$ are employed unless explicitly mentioned.

\section{Basic equations}\label{sec:basic}
\subsection{Hydrodynamics}

The basic equations for the numerical hydrodynamics employed in this work are formulated in the framework of the 3+1 decomposition of the spacetime. In the 3+1 formulation, the metric tensor $g_{\mu\nu}$ is decomposed as
\begin{align}
	ds^2&=g_{\mu\nu}dx^\mu dx^\nu\nonumber\\
	&=-\alpha^2dt^2+\gamma_{ij}\left(dx^i+\beta^idt\right)\left(dx^j+\beta^jdt\right),
\end{align}
where $\mu$ and $\nu$ denote the spacetime indices, $i$ and $j$ the spatial indices, $\alpha$, $\beta^i$, and $\gamma_{ij}$ the lapse, shift, and spatial metric, respectively. We treat the hydrodynamics fluid as a perfect fluid and the energy-momentum tensor is written as
\begin{align}
	T^{\mu\nu}_{\rm fl}=\rho hu^\mu u^\mu+Pg^{\mu\nu},
\end{align}
where $\rho$, $h$, $u^\mu$, and $P$ denote the baryon rest-mass density, specific enthalpy, four-velocity, and pressure, respectively. The equations of energy-momentum conservation and the continuity equation are given by
\begin{align}
	\gamma_{\nu i}\nabla_\mu T^{\mu\nu}_{\rm fl}&=\gamma_{\nu i} G^\nu\label{eq:eom}\\
	n_\nu \nabla_\mu T^{\mu\nu}_{\rm fl}&=n_\nu  G^\nu\label{eq:eoe}\\
	\nabla_\mu \left(\rho u^\mu\right)&=0,\label{eq:eoc}
\end{align}
with the covariant derivative, $\nabla_\mu$. Here, $n_\nu=-\alpha\nabla_\nu t$, $\gamma_{\mu\nu}=g_{\mu\nu}+n_\mu n_\nu$, and $G^\mu$ denotes the radiation four-force density. Equations~\eqref{eq:eom}, \eqref{eq:eoe} and \eqref{eq:eoc} are rewritten in the forms (e.g., Ref.~\cite{2011PThPh.125.1255S})
\begin{align}
	\partial_t S_i&+\partial_k\left(S_i v^k+P\alpha\sqrt{\gamma} \delta^k_i \right)=-S_0\partial_i\alpha\nonumber\\
	&+S_k\partial_i\beta^k-\frac{1}{2}\alpha\sqrt{\gamma}S_{jk}\partial_i\gamma^{jk}+\alpha \sqrt{\gamma}G_i,
\end{align}
\begin{align}
	\partial_t S_0&+\partial_k\left[S_0 v^k+P\sqrt{\gamma} \left(v^k+\beta^k\right) \right]\nonumber\\
	&=-\gamma^{ij}S_i\partial _j \alpha+\alpha\sqrt{\gamma}S_{ij}K^{ij}+\alpha^2\sqrt{\gamma}G^t,
\end{align}
\begin{align}
	\partial_t \rho_*+\partial_k\left(\rho_*v^k\right)=0,
\end{align}
respectively. Here, $K_{ij}$ denotes the extrinsic curvature, and the other variables which newly appear in the above equations are defined as follows:
\begin{align}
\sqrt{\gamma}&={\rm det}\left(\gamma_{ij}\right),\nonumber\\
\rho_*&=\rho w\sqrt{\gamma},\nonumber\\
w&=\alpha u^t,\nonumber\\
S_i&=\rho_* {\hat u}_i=\rho_* h u_i,\nonumber\\
S_0&=\rho_* {\hat e}=\rho_*\left(hw-\frac{P}{\rho w}\right),\nonumber\\
S_{ij}&=\rho hu_iu_j+P\gamma_{ij},\nonumber\\
v^i&=\frac{u^i}{u^t}.\label{eq:varidef}
\end{align}

In our code, we assume axisymmetry of the system. Employing the Cartesian coordinate $(x,y,z)$ and assuming the $z$-axis to be the axis of symmetry,  the system can be describe by the hydrodynamics quantities in the $y=0$ plane. Based on the formulation introduced in Ref.~\cite{2017PhRvD..95h3005S}, the set of hydrodynamics equations, Eqs.~\eqref{eq:eom},~\eqref{eq:eoe}, and~\eqref{eq:eoc}, are rewritten into
\begin{align}
	\partial_t { S}_{x}&+\frac{1}{x}\partial_x\left[x \left( { S}_{x} v^{x}+P\alpha\sqrt{{ \gamma}}\right)\right]+\partial_z\left({ S}_{x} v^{z}\right)\nonumber\\
	&=-{ S}_0\partial_x \alpha+{ S}_{i} \partial_x \beta^{i}-\frac{1}{2}\alpha\sqrt{{ \gamma}}{ S}_{ij}\partial_x { \gamma}^{ij}.\nonumber\\
	&+\frac{1}{x}\alpha\sqrt{{ \gamma}} P +\frac{1}{x}{ S}_y v^y +\alpha \sqrt{\gamma} G_x,\label{eq:eom2x}
\end{align}
\begin{align}
	\partial_t { S}_y&+\frac{1}{x^2}\partial_x\left(x^2{ S}_y v^{x}\right)+\partial_z\left( { S}_y v^z\right)\nonumber\\
	&=\alpha \sqrt{\gamma} G_y,\label{eq:eom2y}
\end{align}
\begin{align}
	\partial_t { S}_z&+\frac{1}{x}\partial_x\left(x { S}_z v^{x}\right)+\partial_z \left( { S}_z v^z+P\alpha\sqrt{{ \gamma}}\right)\nonumber\\
	&=-{ S}_0\partial_z \alpha+{ S}_{i} \partial_z \beta^{i}-\frac{1}{2}\alpha\sqrt{{ \gamma}}{ S}_{ij}\partial_z { \gamma}^{ij}+\alpha \sqrt{\gamma} G_z,\label{eq:eom2z}
\end{align}
\begin{align}
	\partial_t { S}_0&+\frac{1}{x}\partial_x\left\{x\left[{ S}_0v^{x}+P\sqrt{{ \gamma}}\left(v^{x}+\beta^{x}\right)\right]\right\}\nonumber\\
	&+\partial_z\left[{ S}_0v^{z}+P\sqrt{{ \gamma}}\left(v^{z}+\beta^{z}\right)\right]\nonumber\\
	&=-{ \gamma}^{ij}{ S}_{i}\partial_{j} \alpha+\alpha\sqrt{ {\gamma}}{ S}_{ij}K^{ij}+\alpha^2 \sqrt{\gamma} G^t,\label{eq:eoe2}
\end{align}
and
\begin{align}
	\partial_t { \rho}_*+\frac{1}{x}\partial_x\left(x { \rho}_* v^x\right)+\partial_z\left( { \rho}_* v^z\right)=0,\label{eq:eoc2}
\end{align}
respectively. Here, the indices $i$ and $j$ take $x$, $y$, and $z$. 

\subsection{Radiation}
From microscopic point of view, a radiation field is consist of EM waves. If the wavelength of EM waves (the de Broglie wavelength for the case of neutrinos) is much smaller than the typical size of the system, EM (de Broglie) waves can be treated as particles under the geometric optics approximation. In a certain time slice, the state of each photon/neutrino is determined by the spatial coordinates, $x^i$, and the momentum, $p_i$. The evolution of these quantities are determined by the geodesic equations,
\begin{align}
\frac{d x^\mu}{d\lambda}=p^\mu,\,\,\,
\frac{d p_i}{d\lambda}=\Gamma^\mu_{~i\nu}p_\mu p^\nu,
\end{align}
together with the normalization condition,
\begin{align}
g^{\mu\nu}p_\mu p_\nu = -m_{\rm p}^2
\end{align}
where $\lambda$ and $m_{\rm p}$ are the affine parameter and the mass of the particle, respectively. In the following, we set $m_{\rm p}=0$ considering the case of photons or neutrinos (of which mass is negligible compared to the energy scale of the system). By the 3+1 decomposition, these equations can be rewritten in the form suitable to follow the time evolution as~\citep{1994PhRvD..49.4004H,2018MNRAS.475.4186F}
\begin{align}
	\frac{dx^i}{dt}&=\gamma^{ij}\frac{p_j}{p^t}-\beta^i,\label{eq:geo1}\\
	\frac{dp_i}{dt}&=-\alpha\left(\partial_i\alpha\right)p^t+\left(\partial_i \beta^j\right)p_j-\frac{1}{2p^t}\left(\partial_i \gamma^{jk}\right)p_jp_k,\label{eq:geo2}\\
	p^t&=\frac{1}{\alpha}\sqrt{ \gamma^{ij}p_ip_j}.\label{eq:geo3}
\end{align}

To describe a radiation field by the set of photons, it is useful to introduce the distribution function in the phase space. We note that there are at least two options for the coordinates to describe the momentum sector of the phase space: employing the momentum with the upper ($p^i$) or with the lower indices ($p_i$). While the momentum with the upper indices, $p^i$, is also often used for the coordinates of the momentum space (e.g., Ref.~\citep{2014PhRvD..89h4073S}), in our formulation, we employ the momentum with the lower indices, $p_i$, following Refs.~\citep{2009ApJS..184..387D,2015ApJ...807...31R,2018MNRAS.475.4186F} because they match the geodesic equation in the 3+1 form. To avoid the confusion, in the following, we refer to the coordinate volume element of the momentum space with the lower indices, $dp_1dp_2dp_3$, as $d_3p$, while the spatial coordinate volume element, $dx^1dx^2dx^3$, as $d^3x$.

For the phase space described in the coordinates of $(x^\mu,p_i)$, the gauge-invariant phase-space volume element in a time slice is given by $d^3x d_3p=(-n_\mu p^\mu)dV d\Pi$~\citep{1966AnPhy..37..487L,Ehlers1971}, where $dV$ and $d\Pi$ are the spatial volume element and momentum space volume element defined by $dV:=\sqrt{\gamma}d^3x$ and $d\Pi:=d_3p/\left({\sqrt{-g}p^t}\right)$, respectively. We note that, while $d\Pi$ is gauge invariant under spacetime coordinate transformations,  $dV$ is gauge invariant only under spatial coordinate transformations.

The distribution function of photons/neutrinos in the phase space, $f$, is defined by 
\begin{align}
	f (x^\mu,p_i):=\frac{dN}{d^3x d_3p}=\frac{dN}{(-n_\mu p^\mu)dV d\Pi},
\end{align}
where $dN$ denotes the photon/neutrino number in the gauge-invariant phase-space volume $d^3x d_3p=(-n_\mu p^\mu)dV d\Pi$. Note that $f$ is also gauge invariant as $dN$ is gauge invariant quantity.

We can show that $d^3x d_3p=(-n_\mu p^\mu)dV d\Pi$ is also invariant along the geodesic flow (Liouville's Theorem). It follows that the change in the photon/neutrino number per a unit affine parameter in the gauge-invariant phase-space volume is given by
\begin{align}
	\frac{dN}{d^3x d_3pd\lambda}:=\left.\frac{df}{d\lambda}\right|_{\rm source}={\cal L}\left[f\right]\label{eq:bmeq},
\end{align}
where ${\cal L}\left[f\right]$ is the so called Liouville's operator defined by (e.g., Ref.~\citep{2015ApJ...807...31R})
\begin{align}
{\cal L}\left[f\right]:&=\frac{dx^\mu}{d\lambda}\frac{\partial f}{\partial x^\mu}+\frac{dp_i}{d\lambda}\frac{\partial f}{\partial p_i}\nonumber\\
&=\frac{dx^\mu}{d\lambda}\frac{\partial f}{\partial x^\mu}-\frac{1}{2}\left(\partial_i g_{\mu\nu}\right)p^\mu p^\nu\frac{\partial f}{\partial p_i},
\end{align}
and $\displaystyle{\left.\frac{df}{d\lambda}\right|_{\rm source}}$ determines the change in the photon number caused by various radiative process events of photons/neutrinos, such as emission, absorption, and scattering. 

The energy-momentum tensor of a radiation field is given by
\begin{align}
	T^{\mu\nu}_{\rm rad}&=\int d\Pi\,p^\mu p^\nu f=\int \frac{d_3p}{{\sqrt{-g}p^t}} p^\mu p^\nu f.\label{eq:trad}
\end{align}
The conservation law of the total energy-momentum, $\nabla_\mu \left(T^{\mu\nu}_{\rm fl}+T^{\mu \nu}_{\rm rad}\right)=0$, leads to the expression of the radiation four-force density as
\begin{align}
	G^\mu=-\nabla_\mu T^{\mu \nu}_{\rm rad}&=-\int d\Pi\,p^\mu \left.\frac{df}{d\lambda}\right|_{\rm source}\nonumber\\
	&=-\int \frac{d_3p}{{\sqrt{-g}}} p^\mu  \left.\frac{df}{dt}\right|_{\rm source}.\label{eq:ffrad}
\end{align}
Here, $\displaystyle{\left.\frac{df}{dt}\right|_{\rm source}:=\frac{1}{p^t}\left.\frac{df}{d\lambda}\right|_{\rm source}}$ denotes the change in the photon number per a unit time parameter in the gauge-invariant phase-space volume.

\section{Numerical Method}\label{sec:method}
\subsection{Hydrodynamics}
In this work, we solve the set of equations, Eqs.~\eqref{eq:eom2x}--~\eqref{eq:eoc2}, in the conservative form. The numerical flux is calculated by employing a Kurganov-Tadmor scheme~\citep{2000JCoPh.160..241K} with a piecewise parabolic reconstruction for the hydrodynamics quantities of cell interfaces and a steep minmod filter for the flux-limitter. The linear interpolation is used to determine the thermodynamical quantities and four-velocity for each location of the radiation packets. The tetrad frame is constructed from the interpolated four-velocity employing the Gram–Schmidt orthonormalization and used to define the quantities in the fluid rest-frame. The hydrodynamics solver in our code is parallelized by the domain decomposition method with OpenMP.

\subsection{Monte-Carlo scheme for a radiation field}
In the Monte-Carlo scheme that we employ, a radiation field is described by a set of photon/neutrino packets, each of which represents a number of photons/neutrinos. Each packet has information of the position, $x_{(k)}^i$, and the momentum, $p_{(k),i}$ (for the $k$-th packet), which describe the position and momentum of consisting photons/neutrinos. For a given set of packets, the distribution function is approximated by the following form~\citep{2015ApJ...807...31R}:
\begin{align}
&f(t,x^i,p_i)\approx f_{\rm MC}(t,x^i,p_i)\nonumber\\
&:=\sum_k w_{(k)}(t)\delta^3\left[x^i-x_{(k)}^i(t)\right]\delta^3\left[p_i-p_{(k),i}(t)\right].\label{eq:fmc}
\end{align}
Here, $w_{(k)}$ denotes the weight of the packet, which describes how many photons/neutrinos are contained in each packet, and the summation is taken for the packets which are located in a cell coordinate volume of $\Delta^3x$. Substituting the distribution function in Eq.~\eqref{eq:trad} together with Eq.~\eqref{eq:fmc}, the energy-momentum tensor of a radiation field for a given cell of $\Delta^3x$ is expressed as
\begin{align}
	T_{\rm rad}^{\mu\nu}=\frac{1}{\sqrt{-g}\Delta^3 x}\sum_k w_k \frac{p_{(k)}^\mu p_{(k)}^\nu}{p_{(k)}^t}.
\end{align} 

Substituting the distribution function in Eq.~\eqref{eq:bmeq} together with Eq.~\eqref{eq:fmc} and integrating the equation for the infinitesimally small phase volume around the vicinity of each packet with the weights of $x^id^3xd_3p$ and  $p_id^3xd_3p$ lead to the geodesic equations (Eqs.~(\ref{eq:geo1}) and~(\ref{eq:geo2})) for the packet position and momentum ($x_{(k)}^i$ and $p_{(k),i}$). This shows that the evolution of the packet is determined simply by solving the geodesic equation.

The integration for a small phase volume $\Delta^3 x\Delta_3 p$ around the vicinity of each packet gives the evolution equation for $w_{(k)}(t)$:
\begin{align}
	\frac{dw_{(k)}}{dt}=\left.\frac{df}{dt}\right|_{\rm source}\left[x_{(k)}^\mu,p_{(k),i}\right]\Delta^3 x\Delta_3 p.
\end{align}
This implies that the source term of Eq.~\eqref{eq:bmeq} can be described by the change in $w_{(k)}$ or the probabilistic creation/annihilation of packets. 

The change in $w_{(k)}$ or the probabilistic creation/annihilation of packets caused by a radiative process event (emission, absorption, and scattering) induces the back reaction force to the matter field. Each radiative process event that happens in a cell volume $\Delta^3x$ during a time step $\Delta t$ contributes to the four-force density in the following form derived from Eq.~\eqref{eq:ffrad}:
\begin{align}
\Delta G^\mu=-\frac{\Delta p^\mu}{\sqrt{-g}\Delta t \Delta^3 x}.\label{eq:4fden}
\end{align}
Here, $\Delta p^\mu$ denotes the four-momentum change occurred in the radiative process event.

A radiation field is evolved by considering the emission, propagation along geodesics, absorption, and scattering of packets. The detailed implementation of the radiative process events (emission, propagation, absorption, and scattering) in our code are presented below. The radiative-transfer solver in our code is parallelized with OpenMP dividing the packet calculations across individual compute cores.

\subsubsection{emission}\label{sec:method:ems}
At the beginning of the radiation field evolution at each time step, packets are created in each cell in the way similar to Ref.~\cite{2015ApJ...807...31R}. For given cell coordinate volume $\Delta^3 x$ and time interval $\Delta t$, the candidate for the number of packets created in the cell, $N'_p$, is determined by
\begin{align}
	N'_p=\sqrt{-g}\Delta t\Delta^3x\,\frac{\eta}{E_{\rm p}},
\end{align}
where $\eta$ and $E_{\rm p}$ denote the (wavelength-integrated) total emissivity and total fluid rest-frame energy of the created packet, respectively, and $E_{\rm ems}=\sqrt{-g}\Delta t\Delta^3x\,\eta$ denotes the total emitted fluid rest-frame energy in the cell. $E_{\rm p}$ is determined by $1/N_{\rm trg}$ of the radiation energy in the cell in thermal equilibrium, $E_{\rm rad,th}$. $N_{\rm trg}$ is a parameter which approximately controls how many packets are used to resolve the fluid rest-frame radiation energy in thermal equilibrium.

As we explain below, our code employs the higher-order time integration scheme for solving the interaction between matter and radiation fields. To guarantee the energy-momentum conservation of the system, our higher-order time integration scheme requires the number of created packets to be a multiple of 12. For this purpose, for the case of $N'_p\ge6$, we set the number of packets created in the cell, $N_p$, to be $N_p=12 [(N'_p+6)/12]$, where $[(N'_p+6)/12]$ denotes the largest integer smaller than $(N'_p+6)/12$. If $N'_p<6$, we give up applying the higher-order time integration scheme and employ the partially first-order scheme with $N_p=N'_p$ as explained in Appendix~\ref{app:pmix}.

Once $N_p$ is determined, the locations of the created packets are determined randomly following a probability density function proportional to $\Sigma(x,z)=\oint d\varphi x \rho(x^i)$. For this purpose, the spatial dependence of the rest-mass density is considered up to the linear order to ensure the second-order accuracy in space of our code for the emission process.

After the location of the packet is determined, we sample the fluid rest-frame energy of the packet (and hence, the consisting photons/neutrinos in it), $\nu_{(k)}$, following the energy dependence of emissivity. Then, the direction of the momentum for each packet, $p^\mu_{(k),{\rm emitted}}$, is determined by a random sampling from isotropic distribution in the fluid rest-frame. Finally, the packet weight for each packet is determined by $w_{(k)}=E_{\rm ems}/(N_p \nu_{(k)})$ so that the total energy of the created packets, $\sum_{k,{\rm created}} w_{(k)} \nu_{(k)}$, agrees with $E_{\rm ems}$. The back reaction of the emission to the matter filed is determined from Eq.~\eqref{eq:4fden} by setting $\Delta p^\mu=\sum_{k,{\rm created}}w_{(k)}p^\mu_{(k),{\rm emitted}}$.

\subsubsection{free-streaming propagation}
The free-streaming propagation of each packet is described by the geodesic equations (Eqs.~\eqref{eq:geo1} and \eqref{eq:geo2}). While the fixed back ground spacetime is employed in this paper, our code is designed to work in the dynamical spacetime obtained by solving Einstein's equation. For this purpose, we solve the geodesic in the Cartesian coordinates. This is because the evolution of the metric field is often defined so in Einstein solvers (e.g., Ref.~\citep{2012PThPh.127..535S}). In particular, for solving the axisymmetric system, the so-called cartoon method is often employed~\cite{2001IJMPD..10..273A,2000PThPh.104..325S}. Since the metric field is only solved and given in the meridional plan in this method, the packet position and momentum is always rotated around the axis of symmetry after the propagation so that the packet is always located in the meridional plane. Practically, we employ the third-order Runge-Kutta scheme for the time integration and the fourth-order Lagrange scheme for the interpolation of the metric variables. 

\subsubsection{absorption and scattering}\label{sec:absandsct}
In our Monte-Carlo code, the absorption and scattering events are treated probabilistically. To generate a random value, we use the Mersenne Twister implemented in Ref.~\cite{MTURL}. For the evolution of a packet, we first propagate the packet freely along the geodesic for the time interval of the hydrodynamics evolution, $\Delta t$. If the packet crosses the cell boundary during the free-streaming propagation, the interval between the initial time and the time at which the packet crosses the boundary is stored as $\Delta t_{\rm cell}$. Otherwise, a value larger than $\Delta t$ is set to $\Delta t_{\rm cell}$.

Next, we determine the time interval between the initial time and time of the first absorption or scattering event, $\Delta t_{\rm event}$. $\Delta t_{\rm event}$ is determined from $\Delta \tau(\Delta t_{\rm event})$, which denotes the optical depth until the next absorption or scattering event. Because the probability for a packet to evolve without being absorbed or scattered for $\Delta \tau$ is given by ${\rm exp}\left({-\Delta \tau}\right)$, $\Delta \tau$ is probabilistically given by $-{\rm ln}\,r$ with $r$ being a random variable uniformly distributing in $(0,1]$. We determine the function form of $\Delta \tau(\Delta t_{\rm event})$ by a linear interpolated function employing the interaction cross-section at $t$ and $t+\Delta t$ following the method of Ref.~\cite{2009ApJS..184..387D}. Specifically, in our code, we obtain $\Delta t_{\rm event}$ by solving
\begin{align}
	\Delta \tau&=\int_0^{\Delta t_{\rm event}}\frac{d\tau}{dt}(t') dt',
\end{align}
where
\begin{align}
	 \frac{d\tau}{dt}(t')&\approx(1-\Delta(t'))\left.\kappa_{\rm tot} \rho \frac{\nu}{p^t}\right|_{t}+\Delta(t')\left.\kappa_{\rm tot} \rho \frac{\nu}{p^t}\right|_{t+\Delta t},\\
	 \kappa_{\rm tot}&=\kappa_{\rm abs}+\kappa_{\rm sct},\,\Delta(t')=\frac{t'-t}{\Delta t},
\end{align}
and $\kappa_{\rm abs}$ and $\kappa_{\rm sct}$ denote the absorption and scattering opacity, respectively. By this implementation, the second-order accuracy with respect to the spatial discretization is ensured for the optical depth estimation. 

If $\Delta t$ is smaller than $\Delta t_{\rm event}$ and $\Delta t_{\rm cell}$, the evolution of the packet for the current time step is finished. If $\Delta t_{\rm cell}$ is smaller than  $\Delta t$ or $\Delta t_{\rm event}$, the packet is pulled back to the initial state at $t$, and freely propagated again for $\Delta t_{\rm cell}$. If $\Delta t_{\rm event}$ is the smallest among these three time intervals, we pull back the packet to the initial state at $t$, and freely propagate it again for $\Delta t_{\rm event}$. After the propagation, we determine the type of the event by the value of $s$, which is randomly sampled from a uniform distribution of $[0,\kappa_{\rm tot}/\kappa_{\rm abs}]$; if $s<1$, we regard the event as an absorption event, and if not, we regard the event as a scattering event.

If the event is an absorption and the total energy of the packet is sufficiently smaller than the internal energy of the cell in which absorption occurs, we employ a simple approach to describe the absorption event: we annihilate the packet and sum up its contribution to the radiation feedback by Eq.~\eqref{eq:4fden} with $\Delta p^\mu=-p^\mu_{\rm absorbed}$ where $p^\mu_{\rm absorbed}$ is the four-momentum of the absorbed packet. On the other hand, if the total energy of the packet is comparable to or larger than the internal energy in the cell, we employ the continuous absorption method following Ref.~\cite{2009ApJS..184..387D}. In this method, the absorption process is treated as the continuous reduction of the packet weight along the free-streaming propagation with the extinction factor given by ${\rm exp}\left[-\int \kappa_{\rm abs}\rho \nu/p^t dt\right]$. At the same time, we count up the contribution to the radiation feedback force consistent with the lost by the extinction by Eq.~\eqref{eq:4fden} with
\begin{align}
	\Delta p^\mu=-\left\{1-{\rm exp}\left[-\int \kappa_{\rm abs}\rho \nu/p^t dt\right]\right\}p^\mu_{\rm ave},
\end{align}
where $\Delta p^\mu_{\rm ave}$ denotes the mean four-momentum calculated by
\begin{align}
	 p^\mu_{\rm ave}=\frac{1}{\Delta t}\int dt\,p^\mu
\end{align}
along the free-streaming propagation with the time interval of $\Delta t$.

This treatment reduces a Monte-Carlo shot noise in the radiation feedback force which can be induced in the optically thin region. Practically, we apply this prescription if the total energy of the packet is larger than $r_{\rm abs}$ times the internal energy in the cell {\it or} if the weight of the packet is larger than $r_{\rm abs}$ times the initial value assigned at the time of the packet creation. Note that, if the continuous absorption method is applied, the continuous reduction of the packet weight is done for {\it every} free-streaming propagation process, and instead, the absorption coefficient is set to be $0$ for judging the type of the radiative process event in order to avoid the double counting of the absorption effect.

For the case that the event is a scattering event, we determine the new fluid rest-frame energy and momentum of packets based on the property of the scattering. For simplicity, we only consider elastic and isotropic scattering processes in this paper as the first step following Ref.~\cite{2018MNRAS.475.4186F}. By this setting, the fluid rest-frame energy is kept  unchanged during the scattering process, and the direction of the momentum is sampled from an isotropic distribution in the fluid rest-frame. The back reaction of the scattering process to the matter field is determined from Eq.~\eqref{eq:4fden} by setting $\Delta p^\mu=p_{\rm out}^\mu-p_{\rm in}^\mu$ with $p_{\rm in}^\mu$ and $p_{\rm out}^\mu$ being the four-momentum of the packet before and after the scattering event, respectively.

\subsection{Residual packet prescription}\label{sec:res}

For maintaining the consistency of the energy-momentum conservation in numerical computation, we need to take into account all the emission processes in the entire simulation region. However, it is numerically inefficient to assign normal packets to the emission from all the cells because the packets created in the cell with very low emissivity carries only a tiny amount of energy. To reduce the computational cost for solving radiation-transfer equations in such a region, we introduce a new prescription of "the residual packet" described as follows.

If the local emissivity of the cell is smaller than a certain value, $\eta_{\rm min}$, we set a flag of "the residual packet" to all the packets created in the cell. During the evolution of a radiation field, the residual packets are evolved in the same way as for the normal packets. At the end of the radiation-field evolution, the residual packets are collected, and the total laboratory-frame energy, momentum, and packet weight of them are recorded for cell by cell in which the packets were located. At the beginning of the next radiation-field evolution, $N_{\rm res}$ residual packets are again created in the center of each cell so that their total laboratory-frame energy, momentum, and packet weight agree with those recorded in the last step. The photon/neutrino energy of the residual packets is determined to be consistent with the packet energy and weight.

Because the residual packets are collected in each time step, the number of the residual packets is always smaller than $N_{\rm grid}\times N_{\rm res}+N_{\rm res,0}$ with $N_{\rm grid}$ and $N_{\rm res,0}$ being the total grid number and residual packet number created in the current time step, respectively. Hence, by this prescription, we can avoid the accumulated increase of the packet number guaranteeing the total energy-momentum conservation with the accuracy of the geodesic solver and machine precision. Note that, since the information of the energy distribution and higher moments of the angular distribution are lost in this procedure, we should keep in mind that $\eta_{\rm min}$ should be kept sufficiently small so that the radiation feedback from the residual packets is not significant.

\subsection{Implicit Monte-Carlo scheme}
The minimum time step required to stably solve the hydrodynamics evolution and packet propagation is given approximately by the light crossing time scale $\Delta t_{\rm LC}\sim \Delta x$ of the grid cell, if the interaction between matter and radiation is negligible. 
However, for the case that the interaction between matter and radiation becomes important, the time scale of the interaction can be much shorter than $\Delta t_{\rm LC}$. 

For instance, the time scale of the emission and its back reaction to the matter field can be estimated by $\Delta t_{\rm ems}\sim e_{\rm fl}/\eta$ where $e_{\rm fl}$ and $\eta$ are the internal energy density of the fluid in the fluid rest-frame and emissivity, respectively. The ratio between the time scales $\Delta t_{\rm ems}/\Delta t_{\rm LC}\sim(e_{\rm fl}/aT^4)/(\alpha_{\rm abs} \Delta x)$ can be much smaller than unity for the case that the absorption coefficient is large and the temperature is high. In such a situation, the time step required to properly solve the system with explicit methods becomes too small to follow the evolution within realistic computational time. To overcome this problem, implicit schemes are often employed in mesh-based radiation hydrodynamics  solvers~\citep[][]{2014ApJS..214...16N,2014ApJS..213....7J,2016ApJ...818..162O}.

For a Monte-Carlo based radiation hydrodynamics solver, the so-called implicit Monte-Carlo scheme is often employed~\citep{1971JCoPh...8..313F,2012ApJ...755..111A,2015ApJS..217....9R,2018MNRAS.475.4186F,2020ApJ...902L..27F,2022ApJ...933..226R}. In this scheme, the absorption opacity, $\kappa_{\rm abs}$, and the scattering opacity, $\kappa_{\rm sct}$, are modified by introducing a parameter (the Fleck factor), $\alpha_{\rm eff}$, as
\begin{align}
	\kappa'_{\rm abs}=\alpha_{\rm eff}\kappa_{\rm abs},\,\kappa'_{\rm sct}=\kappa_{\rm sct}+(1-\alpha_{\rm eff})\kappa_{\rm abs}
\end{align}
for the case that the absorption/emission time scales are shorter than the evolution time step of hydrodynamics equations. By choosing an appropriate value for $\alpha_{\rm eff}$, the interaction time scale between matter and radiation is effectively lengthened so that the system can be stably evolved with a larger time step, while the energy distribution of photons/neutrinos in thermal equilibrium is kept unchanged. 

In our code, the value for $\alpha_{\rm eff}$ is determined so as to satisfy the following conditions: 
\begin{align}
	\alpha_{\rm eff}\leq \frac{1}{\kappa_{\rm abs}\rho\Delta t'}{\rm log}\left[\frac{e_{\rm rad}-\eta/(\kappa_{\rm abs}\rho)}{e_{\rm rad,th}-\eta/(\kappa_{\rm abs}\rho)}\right]\label{eq:alp_eff},
\end{align}
where $\Delta t'=(\Delta t/u^t)$ denotes the time step measured in the fluid rest-frame, $e_{\rm rad}=T^{tt}_{\rm rad}$, and $e_{\rm rad,th}$ denotes the radiation energy density in the local thermal equilibrium state. This condition is derived so that the updated value for the radiation energy density should not overshoot the value in the thermal equilibrium state: The time evolution of radiation energy density in the fluid rest-frame under the one-zone approximation is given by
\begin{align}
	\frac{d e_{\rm rad}}{d t'}=-\kappa_{\rm abs}\rho e_{\rm rad}+\eta.
\end{align}
If we neglect the time dependence of $\eta$ and $\kappa_{\rm abs}\rho$, radiation energy density after $\Delta t'$ is expressed by
\begin{align}
	e_{\rm rad}\left(t'+\Delta t'\right)=\left[e_{\rm rad}\left(t'\right)-\frac{\eta}{\kappa_{\rm abs}\rho }
\right]e^{-\kappa_{\rm abs}\rho \Delta t'}+\frac{\eta}{\kappa_{\rm abs}\rho}.
\end{align}
The requirement for $e_{\rm rad}\left(t'+\Delta t'\right)$ not to overshoot $e_{\rm rad,th}$ gives the condition of Eq.~\eqref{eq:alp_eff}. 

We emphasize that the condition of Eq.~\eqref{eq:alp_eff} reduces to $\alpha_{\rm eff}\leq \left(\kappa_{\rm abs}\rho\Delta t'\beta\right)^{-1}$, which is the same as that employed in the previous studies~\citep{1971JCoPh...8..313F,2012ApJ...755..111A,2015ApJS..217....9R,2018MNRAS.475.4186F,2020ApJ...902L..27F,2022ApJ...933..226R}, for $\kappa_{\rm abs}\rho\Delta t'\ll1$ and for a large value of $\beta=\left.\frac{\partial e_{\rm rad,th}}{\partial e_{\rm gas,th}}\right|_{\rho}$ with $e_{\rm gas,th}$ being the fluid internal energy density in thermal equilibrium. Compared to the condition employed in the previous studies, our prescription of Eq.~\eqref{eq:alp_eff} has an advantage that, as far as the one-zone approximation is valid, the updated radiation and fluid energy do not overshoot the thermal equilibrium values even if they are initially far from the thermal equilibrium condition. 

We should keep in mind that, however, the overshoot may still happen if the updates in the radiation energy density or fluid thermodynamical property are significant due to non-local radiative transfer or the hydrodynamics evolution. While this is a general issue for the implicit Monte-Carlo method, we may be able to solve this issue by taking into account the intermediate-state changes of matter and radiation in the Runge-Kutta sub-step to determine $\alpha_{\rm eff}$ or by forcing matter and radiation to be in thermal equilibrium in an optically thick regime (see also Sec. 6 in Ref.~\cite{2015ApJS..217....9R} for the implicit treatment of adiabatic heating/cooling). We leave the investigation for such ways to future work.

In the implicit scheme, the scattering process induced by the term $(1-\alpha_{\rm eff})\kappa_{\rm abs}$ which comes from the modified scattering opacity $\kappa'_{\rm sct}$ can be regarded as the absorption process immediately followed by thermal emission. Hence, in this process, the fluid rest-frame energy of the photon in the packet should be re-sampled from the thermal distribution while the total energy of the packet is unchanged. For this purpose, at each scattering event, we sample a random variable, $r$, from a uniform distribution in $[0,\kappa'_{\rm sct}]$, and we re-sample the fluid rest-frame energy of the photon from the thermal distribution if $r$ is larger than $\kappa_{\rm sct}$.

\subsection{Higher-order time integration scheme}\label{secIIIE}

In the Monte-Carlo based hydrodynamics codes developed in the previous studies~\citep{2009ApJS..184..387D,2012ApJ...755..111A,2015ApJS..217....9R,2015ApJ...807...31R,2022ApJ...933..226R}, the hydrodynamics sector and radiation sector (packet propagation) are solved with higher-order time integration schemes, but the interaction between them is taken into account by an operator splitting scheme. In this way, the accuracy of the time integration is reduced to the first-order for the case that the interaction between fluid and radiation is important. Such a low convergence order can not only reduce the accuracy of the computation in regions where the optical depth is large and the emissivity is high, but also be the source of numerical instability. In order to improve the numerical accuracy in such a situation, in this work, we propose an iterative method to achieve the higher-order accuracy in time.

In the Runge-Kutta method, the changes in physical quantities are iteratively calculated for the evolution with appropriate time sub-steps and they are combined to achieve the higher-order accuracy in time. However, unlike the usual grid-based computation method, the Monte-Calro scheme involves the creation and annihilation of packets, and thus, it is not clear what is the appropriate definition of the``amount of change'' in a radiation field between certain time steps. Therefore, in this code, we consider a scheme that does not explicitly require to define ``the amount of change'' of a radiation field.

Let ${\bf u}(t)$ and ${\bf y}(t)$ denote the matter and radiation fields at certain time $t$, respectively. Practically, ${\bf u}(t)$ is a vector consisting of a set of values for conserved quantities at each discretized point of the matter field, and ${\bf y}(t)$ is a vector consisting of a set of values for position and momentum of the packets that represent a radiation field. For given ${\bf u}_0={\bf u}(t)$ and ${\bf y}_0={\bf y}(t)$ at the initial time $t$, the time evolution of matter and radiation fields for a time step $\Delta t$ is carried out in the following manner with our higher-order time integration scheme:
\begin{enumerate}
\item{Solve the packet propagation under the initial matter field ${\bf u}_0$ to obtain the radiation field ${\bf y}_1$ at $t+\Delta t$. The matter field is also evolved for $\Delta t$ to obtain ${\bf u}_1$ with an explicit Eulerian scheme incorporating the radiation four-force obtained in the evolution of the radiation field.}
\item{Solve the packet propagation under the matter field ${\bf u}_1$ to obtain the radiation field ${\bf y}_2$ at $t+\Delta t$. The matter field is then evolved for $\Delta t$ to obtain ${\bf u}_2$, incorporating the radiation feedback obtained in the latest radiation-field evolution.}
\item{Calculate the matter field in the intermediate step, ${\bf u}_*=\frac{1}{2}{\bf u}_0+\frac{1}{4}{\bf u}_1+\frac{1}{4}{\bf u}_2$ from the initial matter field ${\bf u}_0$ and the matter field obtained in the previous sub-steps, ${\bf u}_1$ and ${\bf u}_2$.}
\item{Solve the packet propagation under the matter field ${\bf u}_*$ to obtain the radiation field ${\bf y}_3$ at $t+\Delta t$. Also, as in the previous step, the matter field ${\bf u}_3$ at $t+\Delta t$ is obtained by an explicit Eulerian scheme taking the radiation four-force obtained in the evolution of the radiation field into account.}
\item{The radiation and matter fields in the next time step, ${\bf y}_{\rm new}$ and ${\bf u}_{\rm new}$, are calculated by ${\bf y}_{\rm new}=\frac{1}{6}{\bf y}_1+\frac{1}{6}{\bf y}_2+\frac{2}{3}{\bf y}_3 $ and ${\bf u}_{\rm new}=\frac{1}{6}{\bf u}_1+\frac{1}{6}{\bf u}_2+\frac{2}{3}{\bf u}_3$, respectively.}
\end{enumerate}
In this way, if the time step is sufficiently small, the time integration becomes  second-order accurate in the limit of a large packet number even if the interaction between matter and radiation is strong (see Appendix~\ref{app:hoti} for the proof).

In the Monte-Carlo scheme, a radiation field is represented as a set of packets. There are various ways to construct ${\bf y}_{\rm new}$ by the linear combination of the radiation field obtained in each time sub-step, ${\bf y}_1$,${\bf y}_2$, and ${\bf y}_3$. The simplest way is to combine the packets of each radiation field with their packet weights being multiplied by the coefficients of the linear coupling. However, in this way, the number of packets increase accumulatively as time evolves. Hence, instead of modifying the weights of the consisting packets, we construct the radiation field in the next step by combining the radiation fields obtained in each sub-step with their consisting packets being thinned out so that the total packets number in each radiation field is reduced by the degree of the coefficient of linear coupling. If the number of packets is sufficiently large, the radiation field obtained by this way will be equivalent to that obtained by the linear combination. We note that the thinning of the packets should be done carefully keeping the consistency between the remaining packets and the radiation feedback to the matter field; otherwise conservation of energy and momentum of the system will be violated. The practical way of thinning the packets out implemented in our code, which guarantees conservation of the energy momentum of the system to the precision of the geodesic integrator, is described in Appendix~\ref{app:pmix}. 

In Sec.~\ref{sec:test:conv} below, we also perform the computation employing the operator splitting scheme as a comparison. In the operator splitting scheme, the hydrodynamics sector is first solved for each time step without taking the radiation feedback force into account. Then, the radiation field is evolved employing the updated hydrodynamics variables. Finally, the radiation feedback to the hydrodynamics variables obtained during the update of the radiation field is incorporated. Note that the computational cost for solving a radiation field is reduced by a factor of 3 for the operator splitting scheme compared to the higher-order time integration scheme.

\subsection{Optically thick region in a cell}\label{sec:thmreg}

In the region in which the optical depth with respect to the effective opacity defined by $\kappa_{\rm eff}=\sqrt{\kappa'_{\rm abs}(\kappa'_{\rm abs}+\kappa'_{\rm sct})}$ is much larger than 1, matter and radiation are approximately considered to be in thermal equilibrium. Because the distribution function in thermal equilibrium is trivial, the propagation and creation of packets in such a region can be omitted. Based on this concept, in our code, we apply the following prescription to the cell in which such a thermalized region exists. For simplicity, we describe the method in a one-dimensional system along the $x$-axis for instance, while the extension to multi-dimensional space is straightforward.

Let $l_*$ be the propagation distance in the fluid rest-frame for which a packet is approximately thermalized in the opticaly thick region. Practically, we determine $l_*$ by which the effective optical depth, $\kappa_{\rm eff}\rho l_*$, agrees with a certain critical value, $\tau_{\rm therm}$. Whether a packet is in the thermalized region or not is determined in the laboratory frame under the assumption of the stationary velocity field and spacetime during the packet evolution: The largest absolute change in the $x$-coordinate which a packet can have for a given rest-frame time interval, $\Delta t'$, is given by $(\sqrt{(u^x)^2+g^{xx}}\pm u^x)\Delta t'$, where the sign corresponds to the case that the packet is moving toward the $\pm x$ direction, respectively. Let $l_\pm=(\sqrt{(u^x)^2+g^{xx}}\pm u^x)l_*$ and $x_\pm$ being the $\pm x$ boundary of the cell in the laboratory frame, respectively. If a packet is located in $[x_-+l_-,x_+-l_+]$, it is guaranteed that the packet is located in the region in which the packet needs to propagate at least $l_*$ in the fluid rest-frame to reach the cell boundary. Hence, radiation in the region of $[x_-+l_-,x_+-l_+]$ is expected to be in thermal equilibrium if such a region exists.

At the beginning of the evolution of a radiation field, we judge for each grid cell whether it contains a region of which spatial depth measured from the cell boundary in the fluid rest-frame is twice larger than a critical depth $l_*$. A sufficient condition for this requirement is given by $x_+-x_-\geq 2 (l_++l_-)$ using the laboratory frame coordinate. At the same time, we also require that $\kappa_{\rm abs}\rho \Delta t'>1$, where $\Delta t'=\Delta t/u^t$ is the time step measured in the fluid rest-frame, employing the bare value of absorption opacity to guarantee the thermalization within the time step. If there exists such a region, we apply the following procedure to the cell: 
\begin{enumerate}
\item{All packets in the region of $[x_-+l_-,x_+-l_+]$ are collected (absorbed) at the beginning of the evolution at each time step.}

\item{Following the thermal distribution, the packets are sampled and created in the region of $[x_-+l_-,x_+-l_+]$ except for $[x_-+2l_-,x_+-2l_+]$.}

\item{Packets are created following the local emissivity and propagated as usual except for the region of $[x_-+2l_-,x_+-2l_+]$. Packets are absorbed during the propagation if they reach the region of $[x_-+2l_-,x_+-2l_+]$.}

\item{At the end of the evolution, the packets in the region of $[x_-+l_-,x_+-l_+]$ are collected and the packets in the region are sampled again from the thermal distribution.}
\end{enumerate}

By employing this prescription, the packet density in the cell effectively increases since packets are only present in the thin layer close to the surface of the cell, and hence, the Monte-Carlo shot noise of the energy-momentum transport between the cells is reduced for the fixed number of packets.

\section{Code Test}\label{sec:test}
To validate our numerical code, we perform simulations for several test problems.  While we design our code to be applicable in dynamical spacetime, in this paper, we employ the test problems performed on a fixed background metric, which still enable us to validate our radiation hydrodynamics implementation. Except for the Eddington limit test problem, a flat Minkowski metric is employed as the fixed background. For spherically symmetric problems, numerical computations are practically performed in two-dimensional axisymmetric domain but taking the $x$-axis to be the axis of the axisymmetry (i.e., identifying that the $x$-axis and $z$-axis are equivalent in our formulation) and employing only one grid in the radial direction with the reflective boundary condition. Unless otherwise stated, we set $r_{\rm abs}=1$ (Sec.~\ref{sec:absandsct}), $N_{\rm res}=2$ (Sec.~\ref{sec:res}), and $\tau_{\rm therm}=3$ (Sec.~\ref{sec:thmreg}) for the test simulations, and the time interval of the evolution is determined by $\Delta t=0.5 \Delta x$ with $\Delta x$ being the grid spacing. We note that, by the setting of $\tau_{\rm therm}=3$, the prescription introduce in Sec.~\ref{sec:thmreg} is not switched on in the computation expect for the test problem in Sec.~\ref{sec:test:sh4} since the optical depth of each cell is not large.

\subsection{One-zone thermalization}\label{sec:test:therm}

\subsubsection{Energy-independent opacity case}
\begin{figure*}
 	 \includegraphics[width=.32\linewidth]{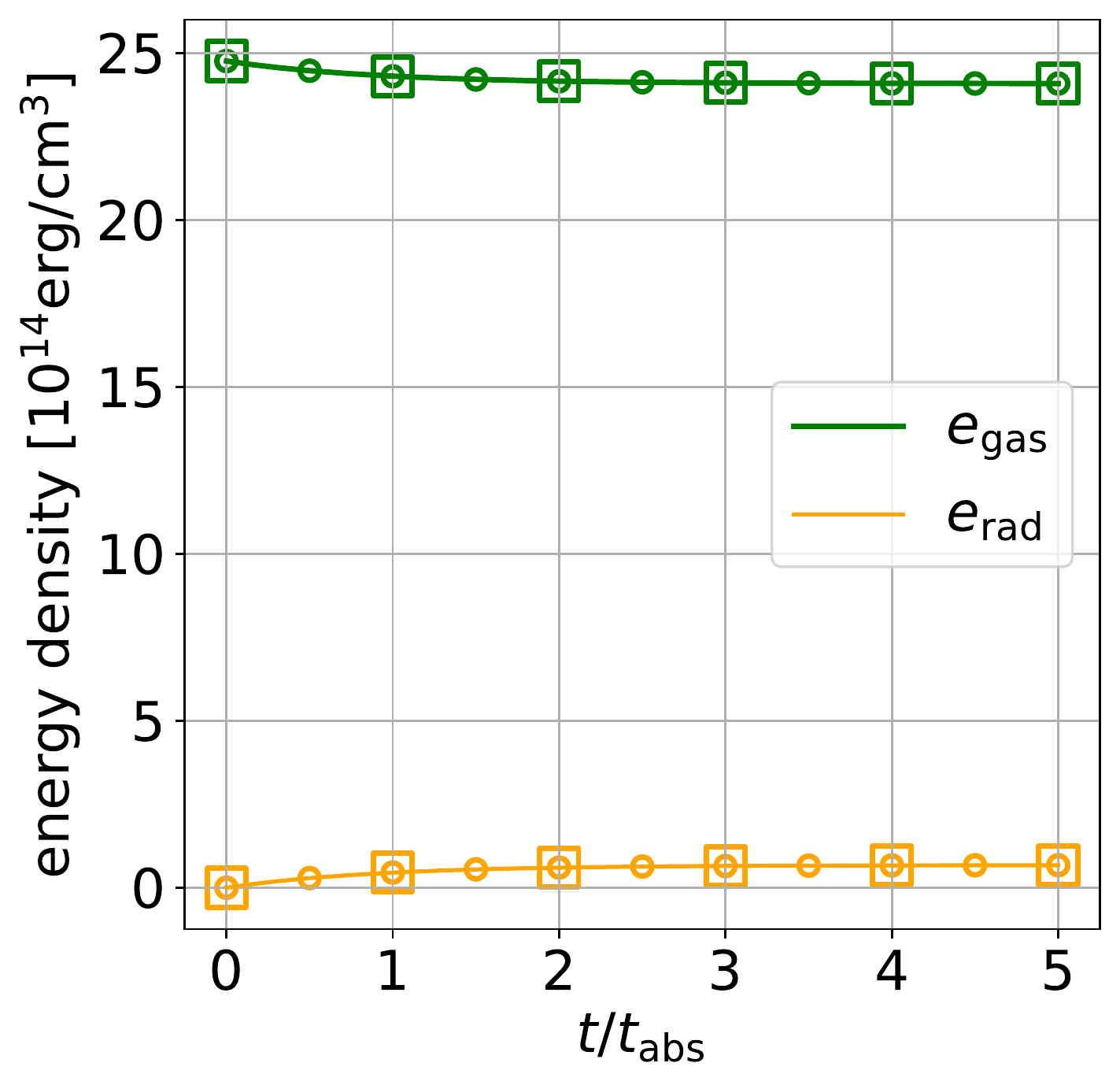}
 	 \includegraphics[width=.32\linewidth]{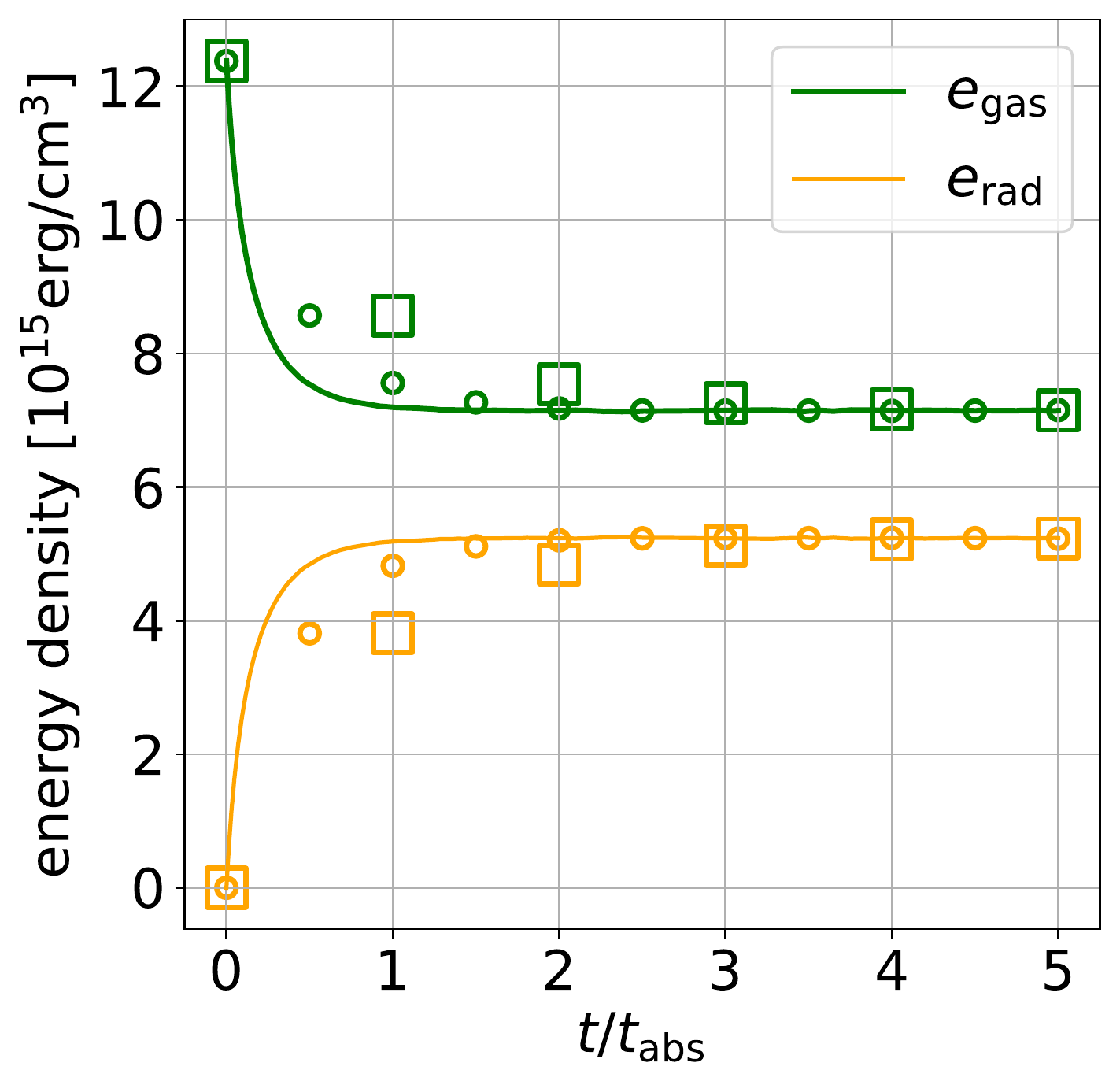}
 	 \includegraphics[width=.32\linewidth]{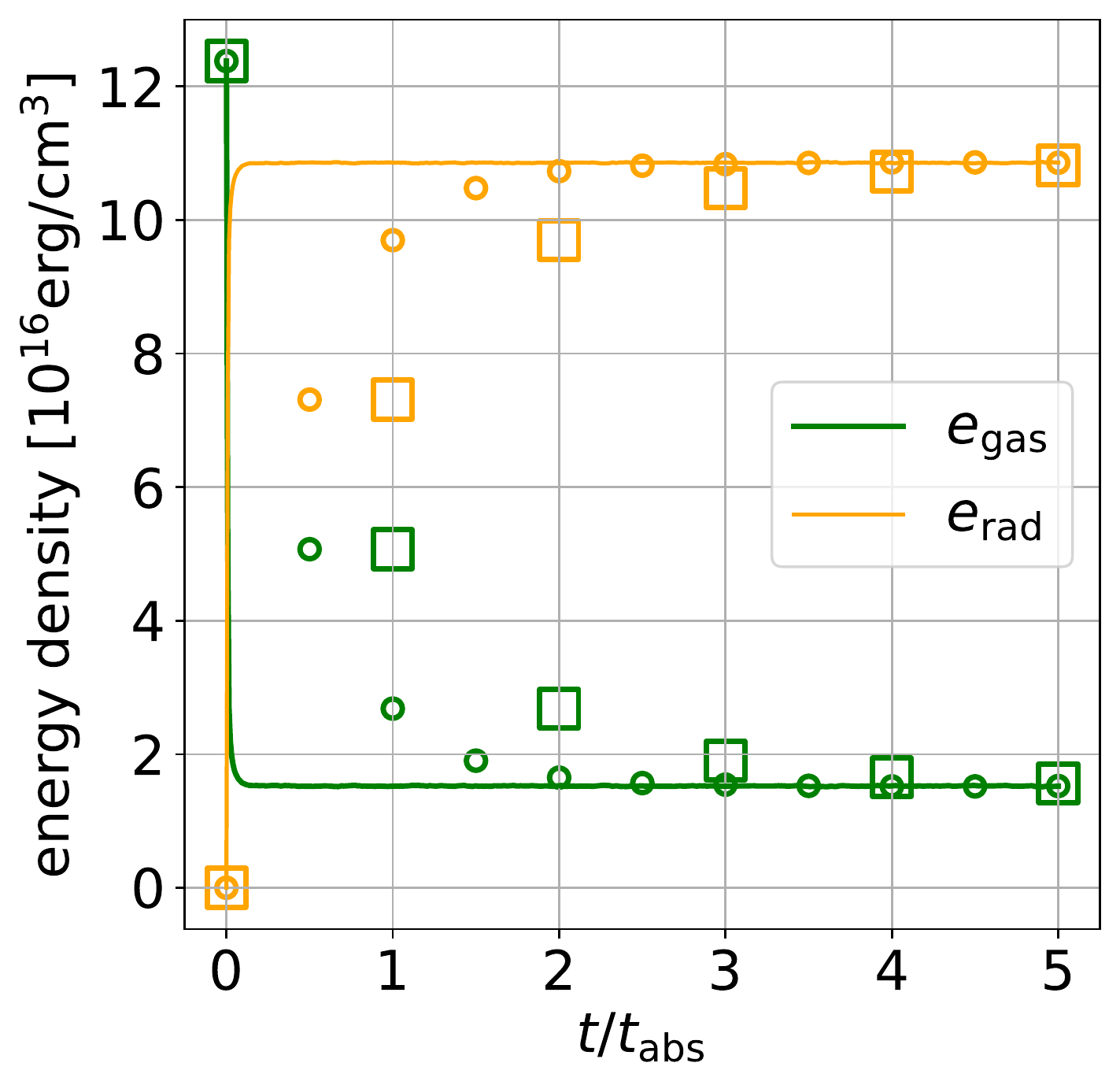}\\
 	 \includegraphics[width=.32\linewidth]{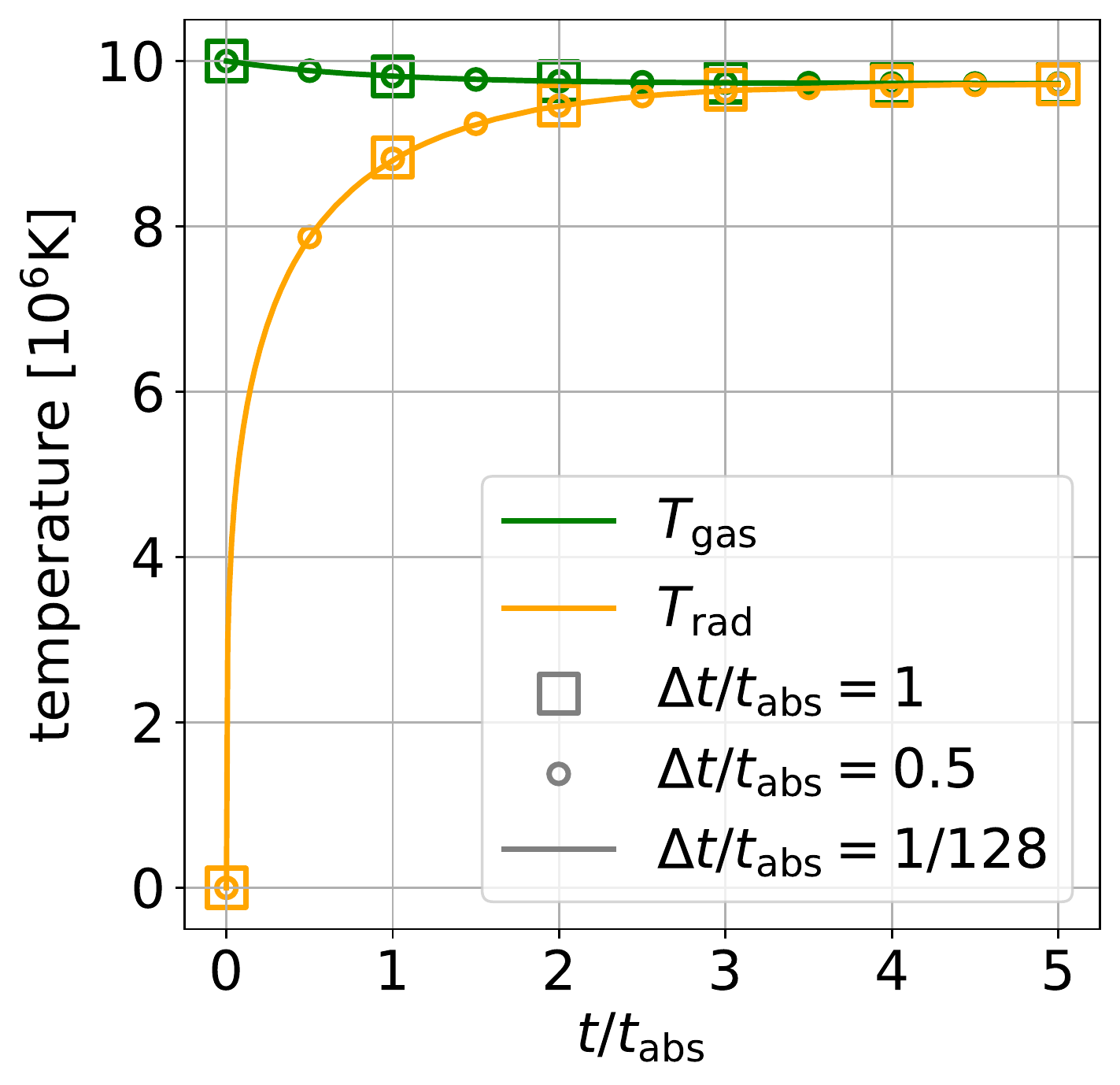}
 	 \includegraphics[width=.32\linewidth]{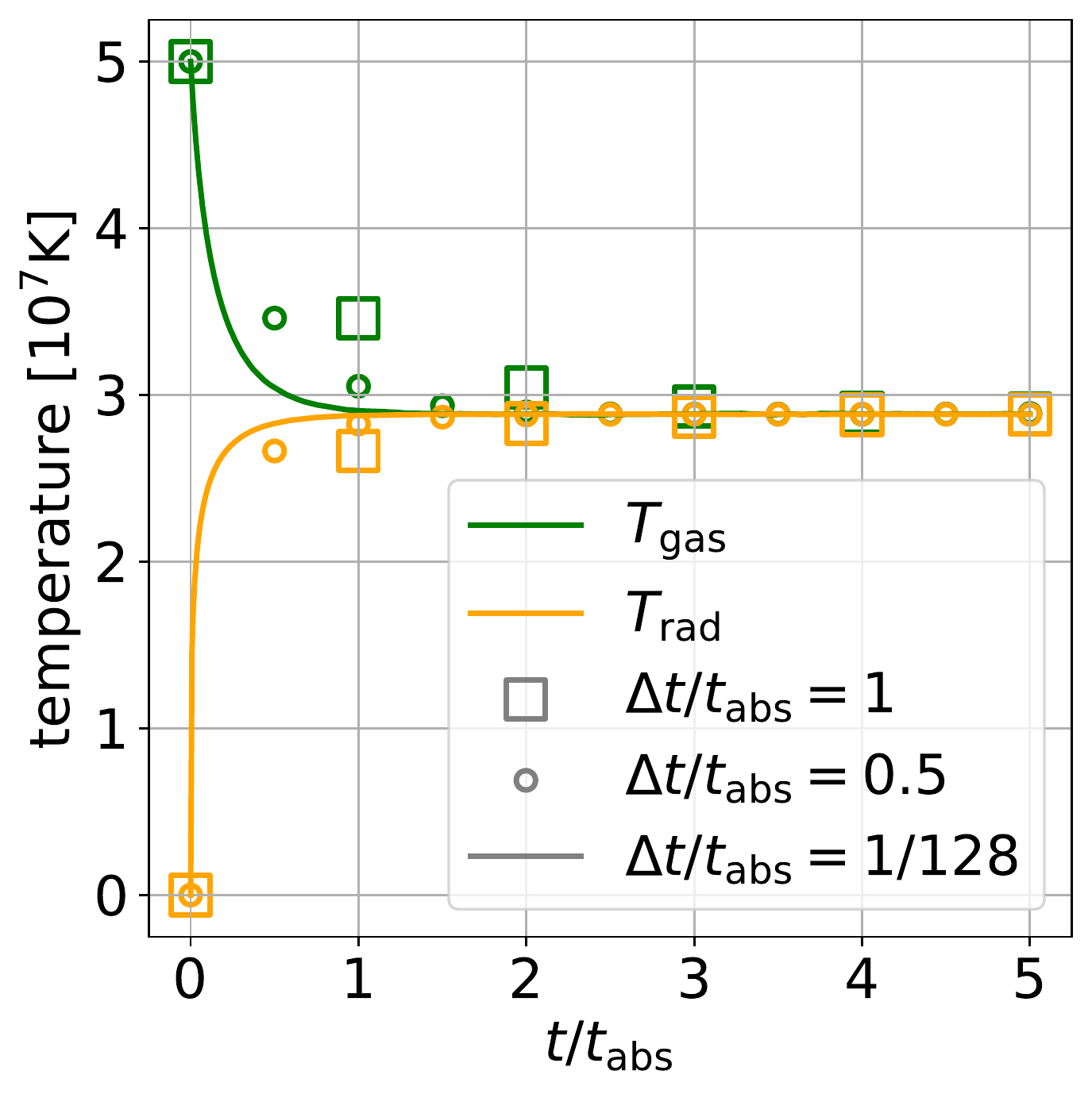}
 	 \includegraphics[width=.32\linewidth]{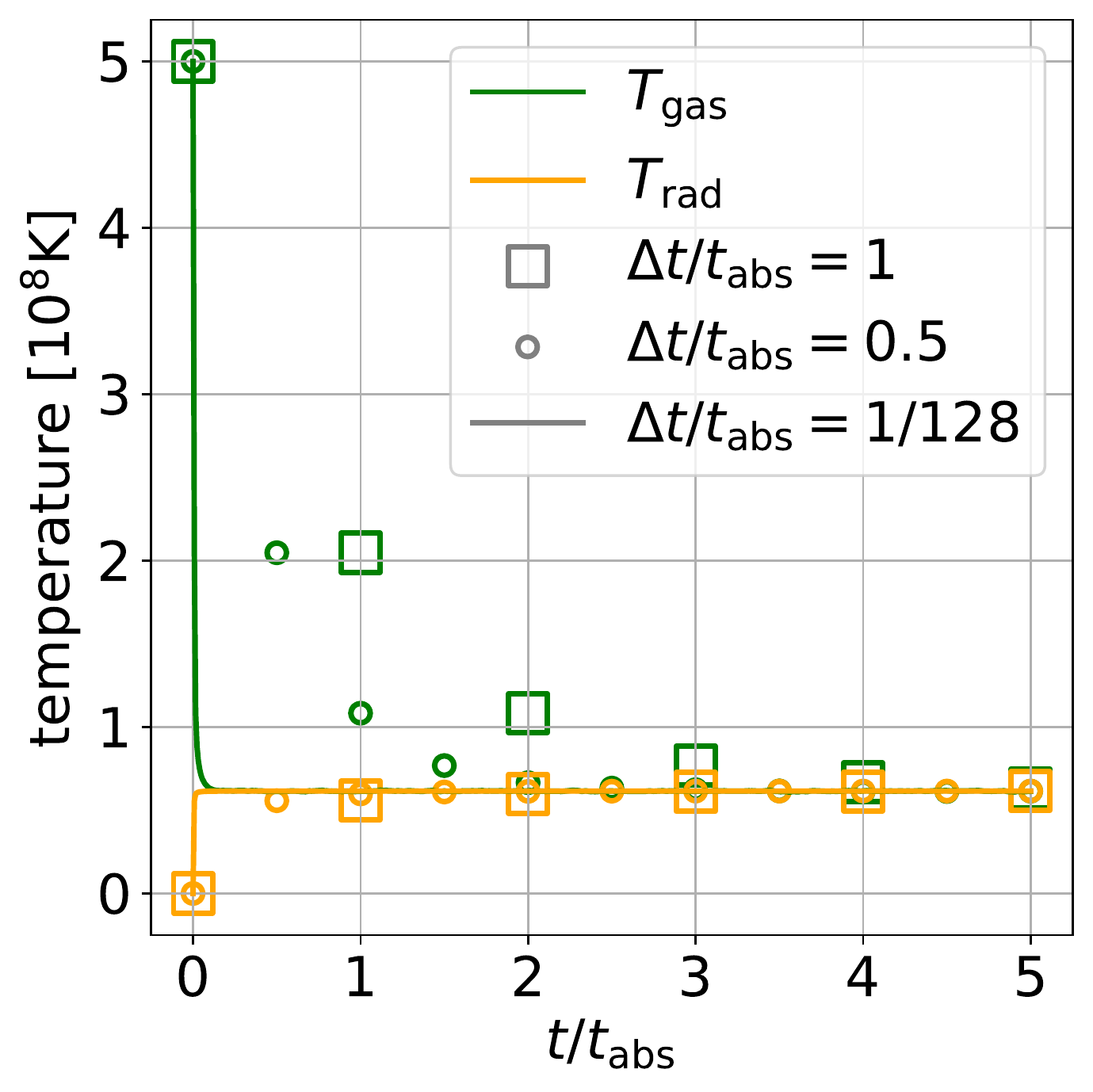}
 	 \caption{Results of the thermalization tests with gray opacity. The left, middle, and right panels are for the cases for which fluid energy is dominant, fluid and radiation energy is comparable, and radiation energy is dominant in the thermal equilibrium states eventually settled, respectively (see the body text for the detailed setups). The upper panels show the fluid internal energy density and radiation energy density. The lower panels show the evolution of the fluid and radiation temperature. The horizontal axes denote the time normalized by the absorption time scale ($t_{\rm abs}=(\kappa_{\rm abs}\rho c)^{-1}$). }
	 \label{fig:therm}
\end{figure*}

To demonstrate that our code can correctly solve the interaction between matter and radiation fields, we evolve homogeneous one-zone systems initially with no radiation but with fluid internal energy, and check whether the systems relax to thermal equilibrium states. In this test, the hydrodynamics grid that consists of a single cell is prepared. We employ the $\Gamma$-law ideal equation of state,  
\begin{align}
	P&=\left(\Gamma_{\rm th}-1\right)e_{\rm gas}, \label{GEOS}\\
	e_{\rm gas}&=\frac{1}{\Gamma_{\rm th}-1}\frac{\rho}{\mu_{\rm ave} m_{\rm u}} k_{\rm B} T_{\rm gas}
\end{align}
 where $T_{\rm gas}$ is the fluid temperature, $\Gamma_{\rm th}=5/3$ is the adiabatic index, $\mu_{\rm ave}=0.5$ is the average molecular weight, and $m_{\rm u}$ is the proton mass. For the rest-mass density, gray absorption opacity, and gray scattering opacity, we employ the values of $\rho=1\,{\rm g\,cm^{-3}}$, $\kappa_{\rm abs}=1\,{\rm cm^2 g^{-1}}$, and $\kappa_{\rm sct}=0$, respectively.  The monochromatic emissivity, $\eta_\nu$, is given by Kirhichoff’s law assuming the Planckian black-body source function as follows:
 \begin{align}
	\eta_\nu&=\kappa_{\rm abs}\rho c\,B_{\nu}\left(T_{\rm gas}\right),\nonumber\\
	B_{\nu}\left(T_{\rm gas}\right)&=\frac{1}{\pi^2\hbar^3c^3}\frac{\nu^3}{e^{\nu/k_{\rm B} T_{\rm gas}}-1}. 
\end{align}
In this test, 3 cases with different initial fluid temperature are examined: $T_{\rm gas}=1\times 10^7$, $5\times 10^7$, and $5\times 10^8\,{\rm K}$, which correspond to the cases that the fluid energy is dominant, fluid and radiation energy is comparable, and radiation energy is dominant in the thermal equilibrium states eventually settled, respectively. For each case of initial fluid temperature, the computation with 3 different time steps ($\Delta t/t_{\rm abs}=$1, 0.5, and 1/128 with $t_{\rm abs}$ being the absorption time scale, $(\kappa_{\rm abs}\rho c)^{-1}$) are performed. For all the computations, we set $N_{\rm trg}=1.2\times 10^5$.
 
Figure~\ref{fig:therm} displays the results of the one-zone thermalization tests. The upper panels show the evolution of fluid internal energy and radiation energy density. The lower panels show the evolution of the fluid temperature and radiation temperature indicator, which is defined by $T_{\rm rad}=(e_{\rm rad}/a_{\rm rad})^{1/4}$ with $a_{\rm rad}$ being the radiation constant. For all the cases, the fluid temperature and radiation temperature relax to an identical value by $t=5\,t_{\rm abs}$, demonstrating that the thermal equilibrium states are achieved.

The time scale for the thermalization depends on the initial internal fluid energy. In particular, this time scale is much shorter than the absorption time scale for the radiation-dominant case; see the result of $\Delta t/t_{\rm abs}=1/128$ in the right panels in Fig.~\ref{fig:therm}. The reason for this is that for this one-zone system, the thermalization time scale is determined basically by the shorter of the absorption time scale ($\approx (\kappa_{\rm abs}\rho c)^{-1}$) or the emission time scale ($\approx e_{\rm gas}/\eta$). Since the total emissivity is proportional to $T_{\rm gas}^4$, the emission time scale becomes much shorter than the absorption time scale for the radiation-dominant case. 

The same temperature is reached after the thermalization regardless of the chosen time-step interval as long as the stable numerical computation is feasible. We find that this is in particular the case for the radiation-dominant case, in which the time-step interval can be chosen to be much longer than the time scale of the emission. This is accomplished by the implicit Monte-Carlo scheme. In fact, without the implicit Monte-Carlo scheme, we find that the system is no-longer stably solved for the radiation-dominant case with the large time step interval ($\Delta t/t_{\rm abs}\sim 1$). However, we should note that applying the implicit Monte-Carlo scheme can cause an artificial delay in the thermalization process due to the limiting of the local temperature change in a single time step. Indeed, we find that it takes $\approx 3\,t_{\rm abs}$ for the fluid and radiation temperature to agree with each other for the largest time-step case in the radiation-dominant test. We might need to keep in mind that this delay might give some artifact in the computation, particularly for the case that a sudden physical temperature change occurs in the system.

\subsubsection{Energy-dependent opacity case}

\begin{figure*}
 	 \includegraphics[width=.49\linewidth]{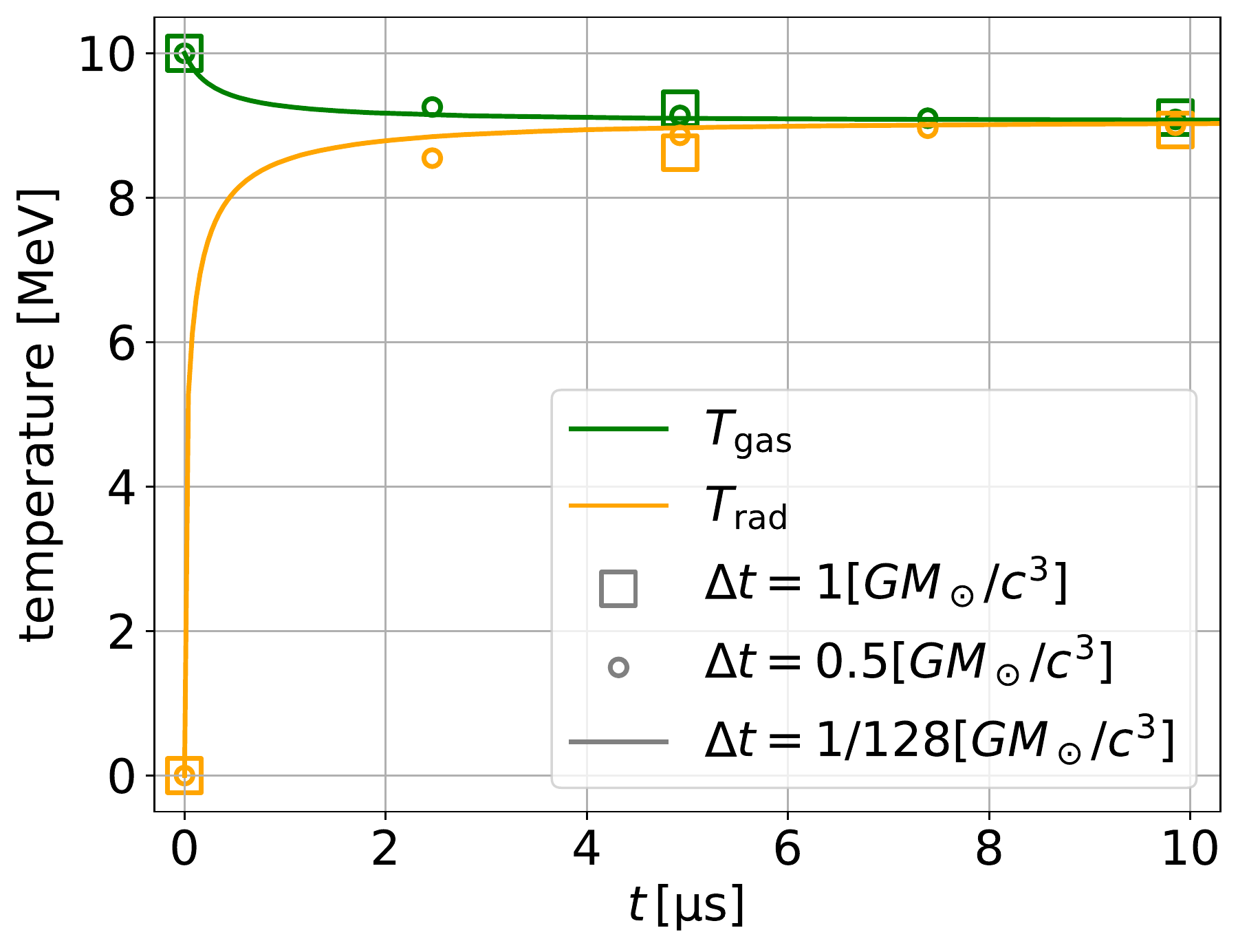}
 	 \includegraphics[width=.49\linewidth]{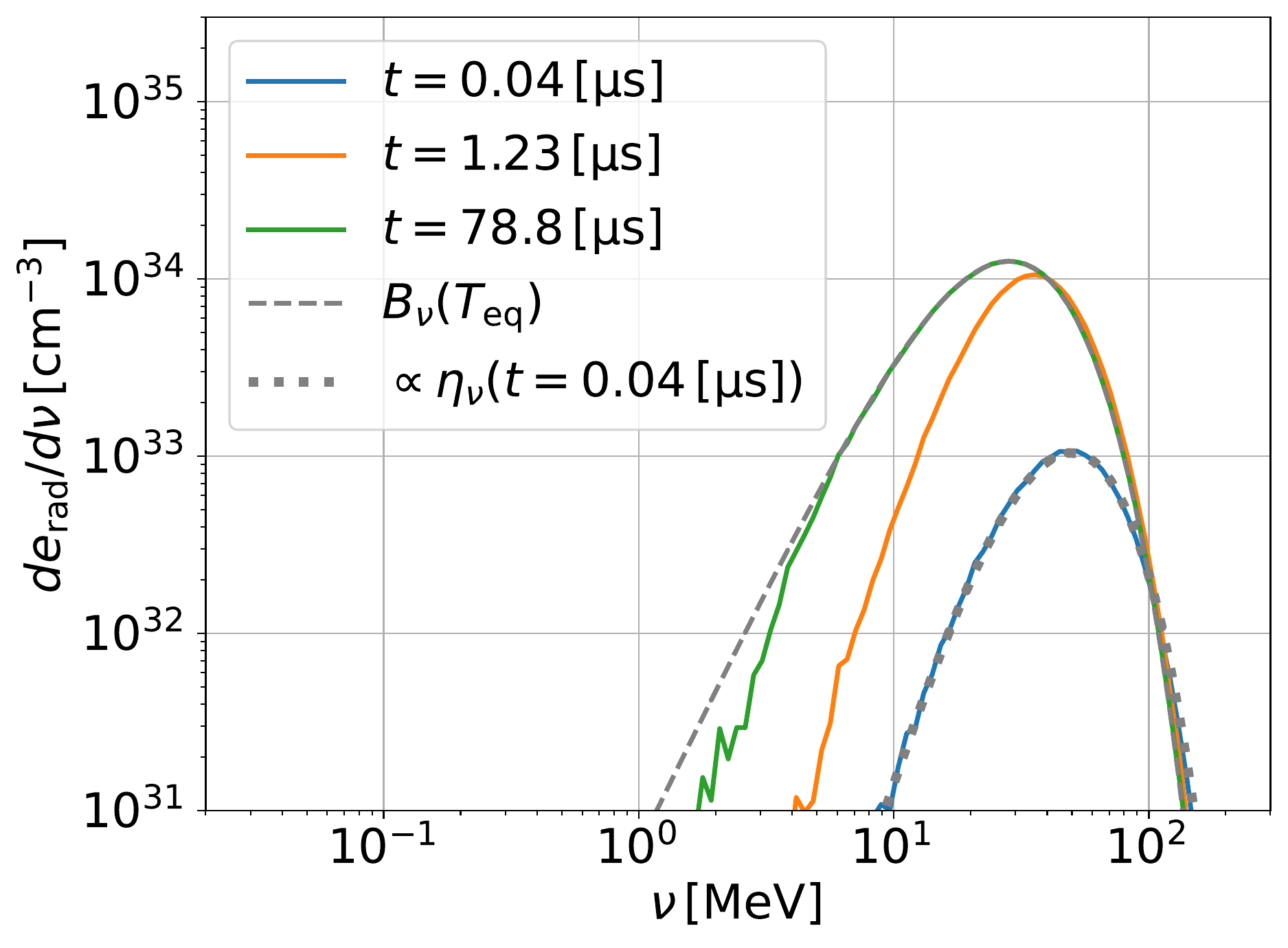}
 	 \caption{Results of the thermalization tests with energy-dependent opacity which mimics neutrino weak interaction. The left panel shows the evolution of fluid and radiation temperature. The right panel shows the energy distribution of radiation at $t=$0.04, 1.23, and 78.8\,${\rm \mu s}$.}
	 \label{fig:therm_nu}
\end{figure*}

In the previous one-zone test, we demonstrated that the thermal equilibrium is appropriately achieved in our code for the fluid with a gray opacity. Next, we examine the similar thermalization test with the energy-dependent opacity. 

For this test, we employ the following setup of opacity and emissivity, mimicing weak interaction between the fluid and neutrino radiation fields:
\begin{align}
	\kappa_{\rm abs}(\nu)&=\frac{G_{\rm F}^2}{m_{\rm u}} \nu^2,\,\kappa_{\rm sct}=0,\nonumber\\
	\eta_\nu&=\kappa_{\rm abs}(\nu)\rho c\,B_{\nu}\left(T_{\rm gas}\right),\nonumber\\
	B_{\nu}\left(T_{\rm gas}\right)&=\frac{1}{\pi^2\hbar^3c^3}\frac{\nu^3}{e^{\nu/k_{\rm B} T_{\rm gas}}+1},
\end{align}
where $\nu$ is the fluid rest-frame energy of neutrino-like particle and $G^2_{\rm F}$ is the Fermi interaction constant given by $5.94\times 10^{-44}\,{\rm MeV^{-2} cm^2}$. We again employ the $\Gamma$-law ideal equation of state with $\Gamma_{\rm th}=5/3$ and $\mu_{\rm ave}=0.5$ for simplicity.  The rest-mass density and the initial fluid temperature are given by $\rho=3\times10^{11}\,{\rm g\,cm^{-3}}$ and $T_{\rm gas}=10\,{\rm MeV}$, respectively, and initially neutrino radiation is set to be absent. We perform the computation with 3 different time step intervals which include the one longer than, one comparable to, and one much shorter than the thermalization time scale. For all the cases, we set $N_{\rm trg}=1.2\times 10^5$.

Figure~\ref{fig:therm_nu} shows the result of the thermalization test. As in the gray opacity case, the fluid temperature and radiation temperature relax to equilibrium values  ($T_{\rm eq}$) as the system evolves regardless of the employed time step interval (see the left panel of Fig.~\ref{fig:therm_nu}). 

The right panel of Fig.~\ref{fig:therm_nu} shows that the energy distribution of radiation agrees approximately with the thermal distribution after the thermal equilibrium is achieved. We also plot the energy distribution of radiation on several time slices obtained with $\Delta t=1/128 [GM_\odot/c^3]$. At the beginning of the simulation ($t=$0.04$\,{\rm \mu s}$), the system has not yet settled into the thermal equilibrium. Indeed, the energy distribution of radiation exhibits a shape close to the emissivity function at that time. As the time evolves, the energy distribution of the radiation field approaches the thermal distribution from the high energy side through interaction between the fluid ($t=$1.23$\,{\rm \mu s}$), and finally, the energy distribution of the radiation field agrees approximately with $B_{\nu}(T_{\rm eq})$ ($t=$78.8$\,{\rm \mu s}$). Note that the low-energy part of the distribution has not yet settled into the thermal equilibrium simply because the interaction time scale $(\kappa_{\rm abs}\rho  c)^{-1}\propto \nu^2$ is longer for lower-energy neutrino-like particle (thus the result is physically reasonable). These results indicate that the thermal equilibrium state is properly achieved in our code even if the opacity has energy dependence. 

\subsection{Test problems from Asahina et al. 2020}
To further validate our code quantitatively, we perform simulations for the same test problems as those examined in Refs.~\cite{2016ApJ...818..162O,2020ApJ...901...96A} (see also the references therein). For all the test problems in this subsection, the same equation of state and Planckian black-body emissivity with a gray opacity as those used in Sec.~\ref{sec:test:therm} are employed but in a system of units in which $e_{\rm gas} =\rho T_{\rm gas}/(\Gamma_{\rm th}-1)$ holds.

\subsubsection{Dynamical diffusion}
\begin{figure*}
 	 \includegraphics[width=.49\linewidth]{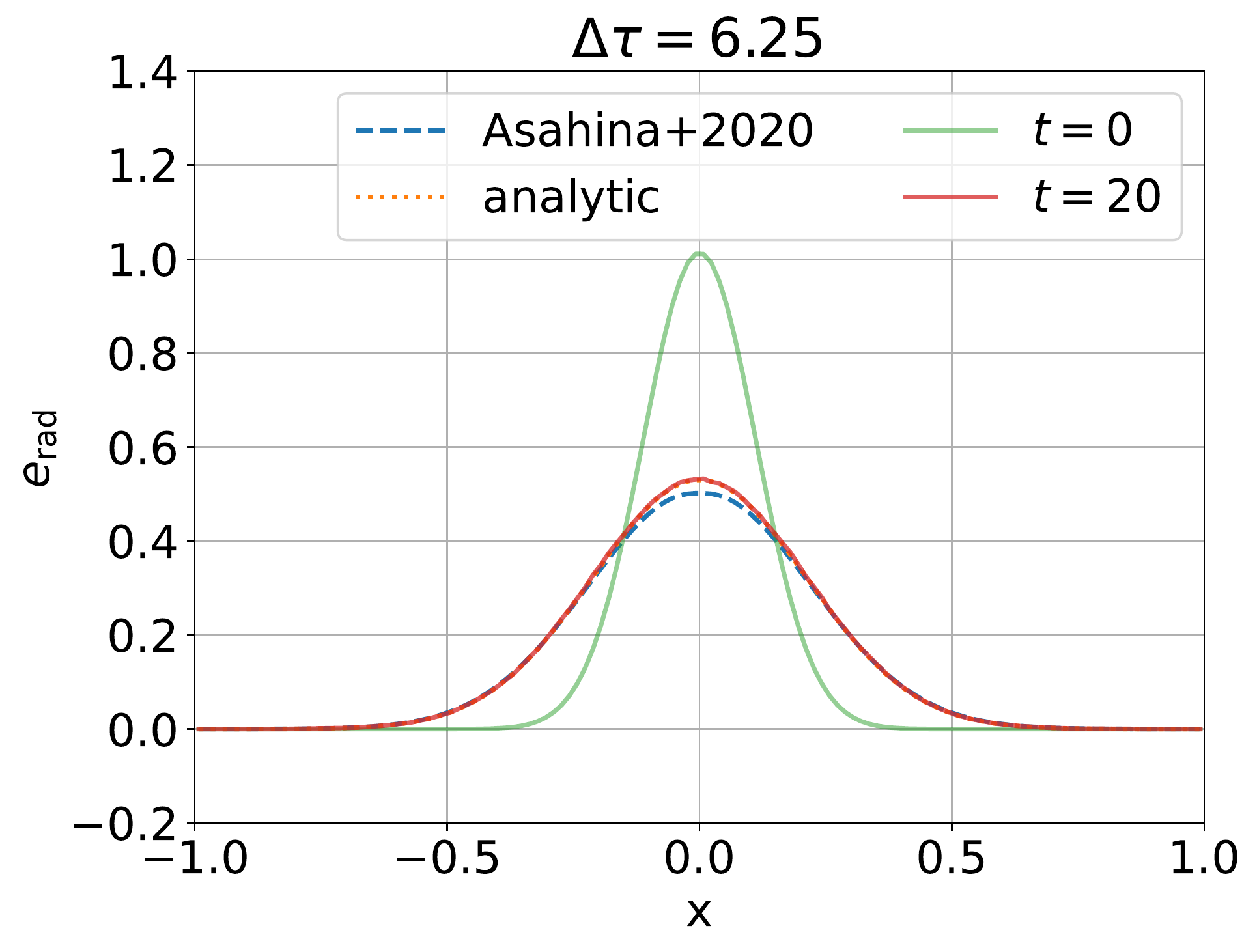}
 	 \includegraphics[width=.49\linewidth]{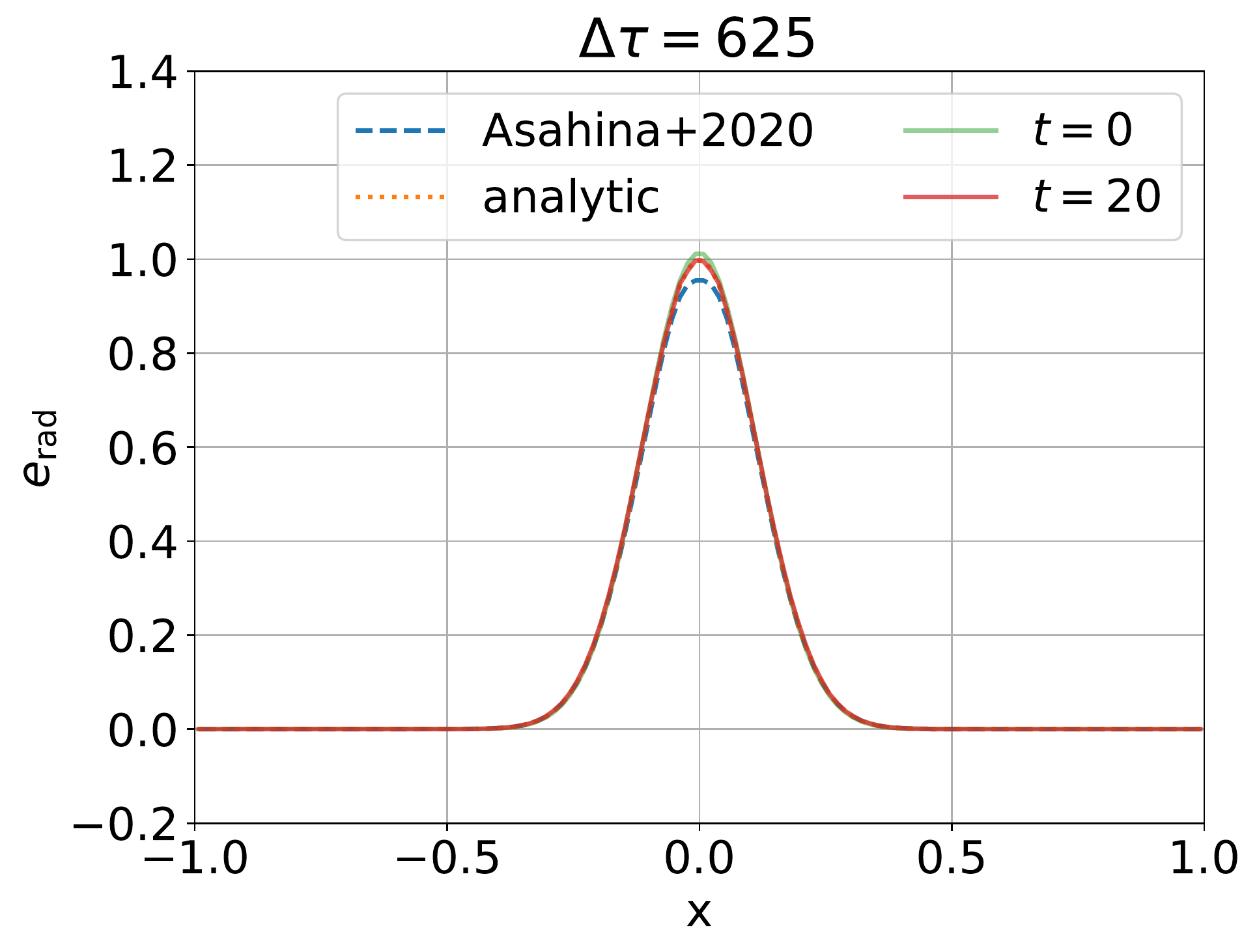}
 	 \caption{Radiation energy density profiles at $t=0$ and 20 for the dynamical diffusion tests. The left and right panels show the results for the test with the scattering optical depth of each cell, 6.25 and 625, respectively. 
 	 The dashed and dotted curves show the results by Asahina et al.~\cite{2020ApJ...901...96A} and analytic solution, respectively.}
	 \label{fig:ddiff}
\end{figure*}
First, we numerically solve the dynamical diffusion test problems which were originally proposed in Ref.~\cite{2014ApJS..213....7J} and employed for the code test in Refs.~\cite{2016ApJ...818..162O,2020ApJ...901...96A} . In this test, the diffusion of radiation in the one-dimensional (1D) homogeneous medium moving toward $+x$-direction with $v^x=0.1\,c$ is considered. The simulation region is prepared as $-1\leq x\leq1$ with 128 grid points, and the periodic boundary condition is applied to the $\pm x$ boundaries. Only the isotropic scattering process is considered, and absorption and emission are not considered. The hydrodynamics variables are fixed during the evolution, and only the evolution of the radiation field is considered in this test. Two different scattering opacity, in which the scattering optical depth of each cell, $\Delta \tau$, becomes 6.25 and 625, is considered in this test. The radiation field is initially set to be isotropic in the fluid rest-frame with the spatial distribution of
\begin{align}
	\left.e_{\rm rad,analytic}(x)\right|_{t=0}={\rm Max}\left[{\rm  exp}\left(-40 x^2\right),{\rm  exp}\left(-10\right)\right].
\end{align}
We always set the number of packets to be $\approx 4\times 10^6$ in this test.

Figure~\ref{fig:ddiff} shows the radiation energy density profiles for the dynamical diffusion tests. The radiation energy is transported following the fluid motion while being diffused, and the peak of distribution comes back to the origin at $t=20$ due to the periodic boundary condition. For both setups of the scattering opacity, the results of our code well  reproduce the analytical solutions. Indeed, we find that the deviation of the numerical results from the analytic ones defined by
\begin{align}
 	\sqrt{\frac{\int^{1}_{-1}\left|e_{\rm rad}(x)-e_{\rm rad,analytic}(x)\right|^2 dx}{\int^{1}_{-1}e_{\rm rad,analytic}(x)^2 dx}}\label{eq:l2error1}
\end{align}
is less than $1\%$ for both $\Delta \tau$ cases at $t=20$.

\subsubsection{Radiation dragging}\label{sec:test:rdrag}
\begin{figure*}
 	 \includegraphics[width=.329\linewidth]{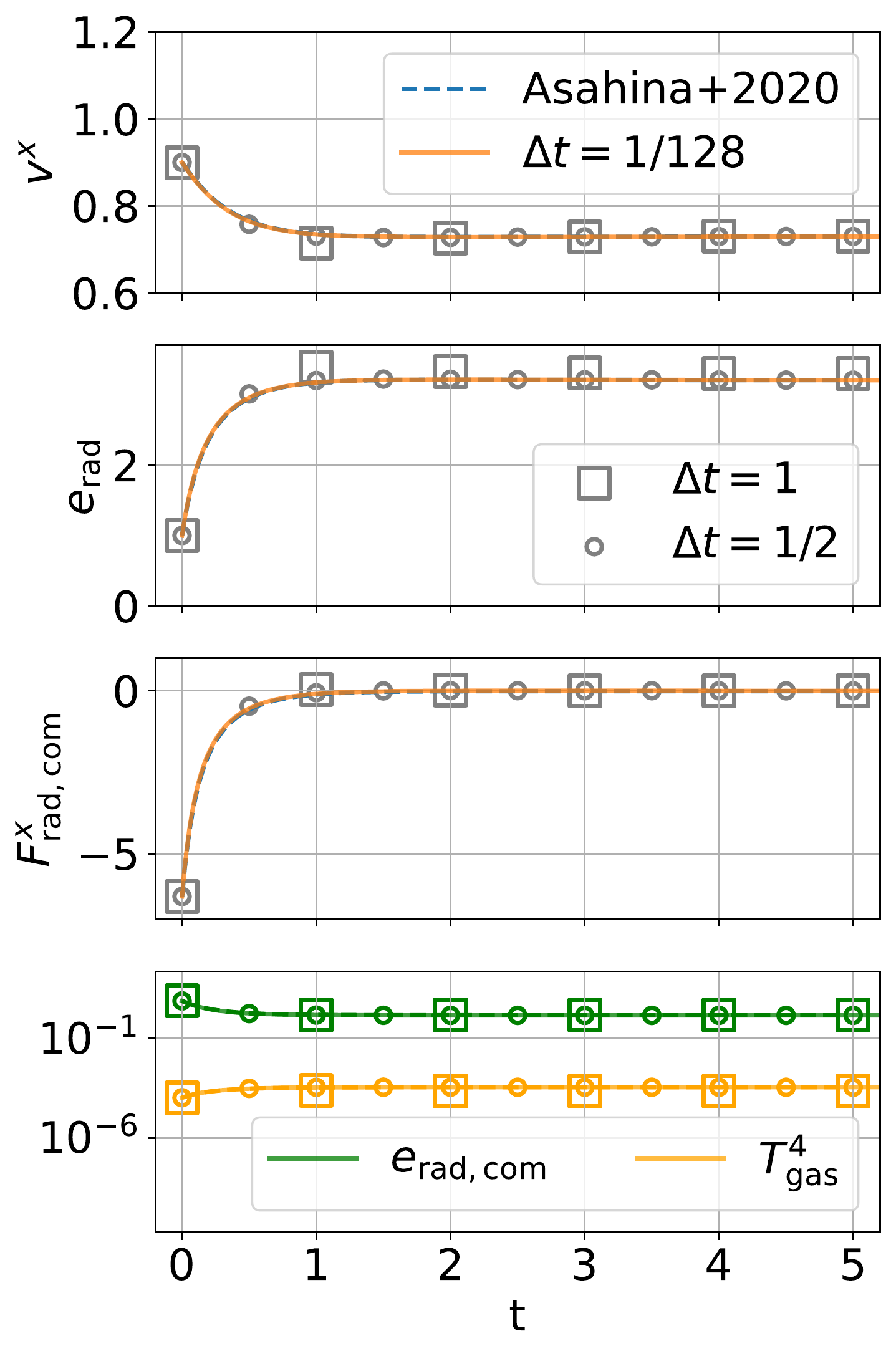}
 	 \includegraphics[width=.329\linewidth]{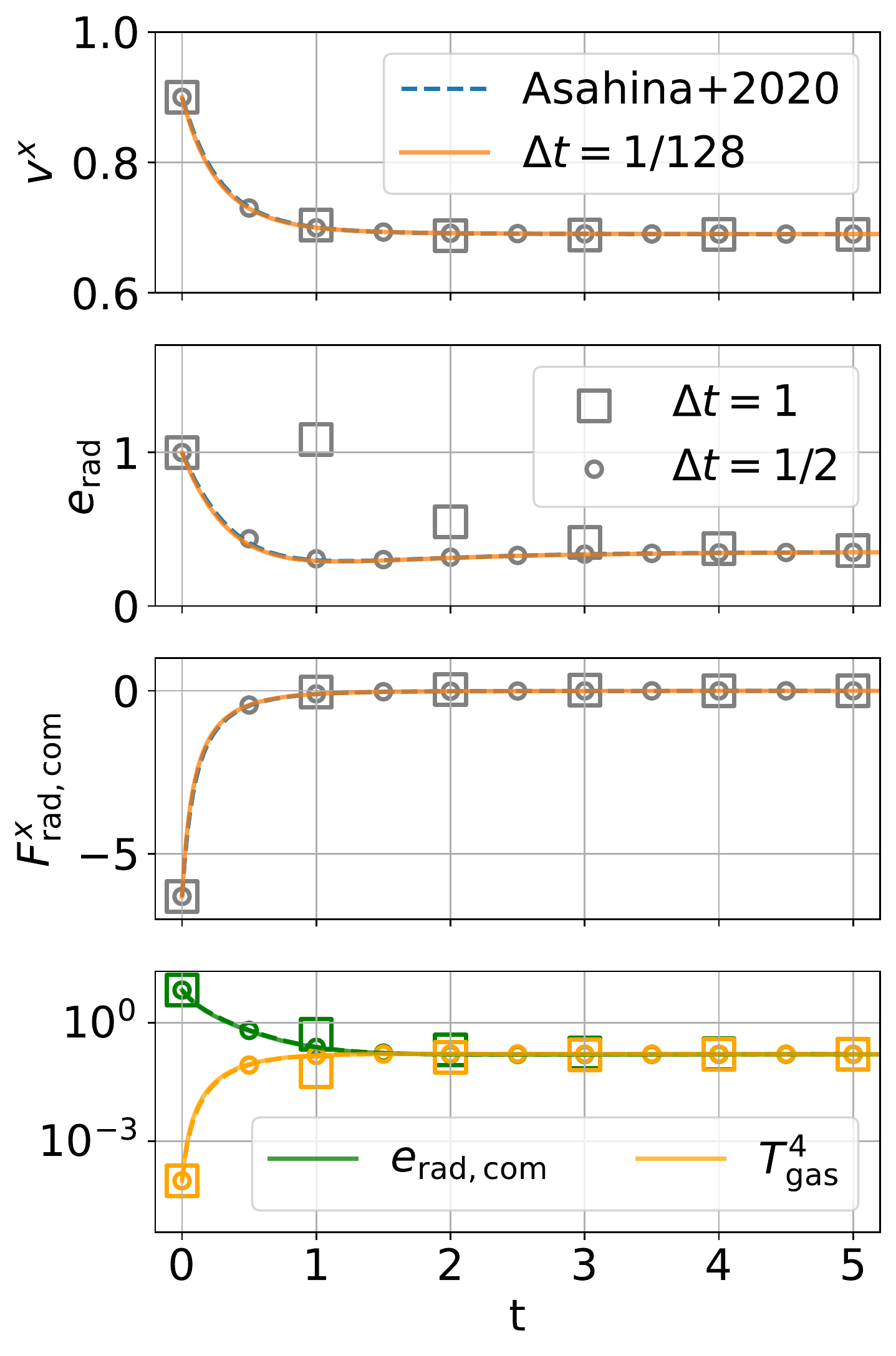}
 	 \includegraphics[width=.329\linewidth]{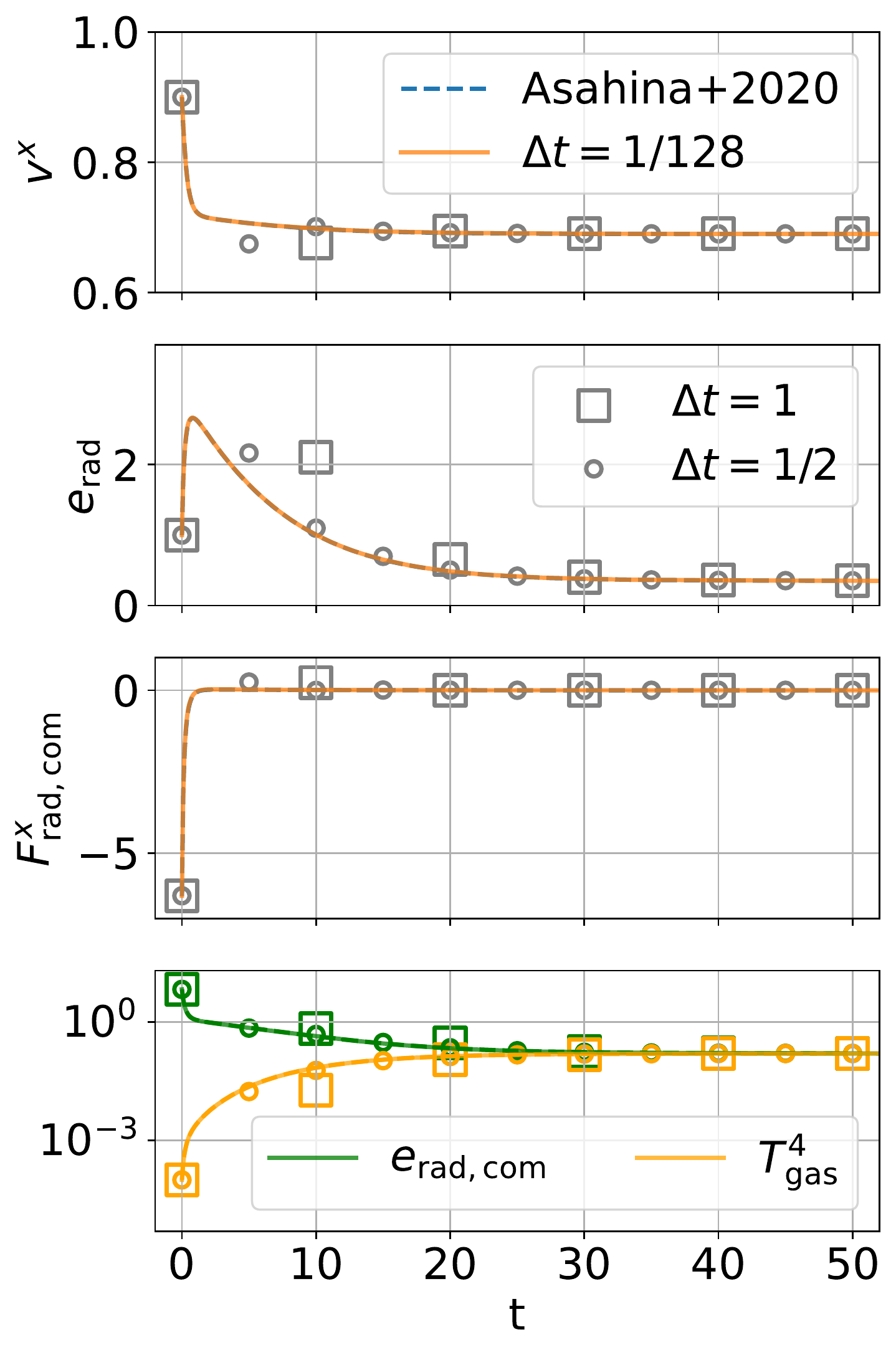}
 	 \caption{Results of the radiation-dragging test problems. The left, middle, and right panels show the results for $(\kappa_{\rm abs}, \kappa_{\rm sct})=(0.0,1.0)$, $(1.0,0.0)$, and $(0.1,0.9)$, respectively. For each case, the $x$-component of fluid velocity, radiation energy density in the laboratory frame, the $x$-component of the radiation flux in the fluid rest-frame, and fluid and radiation energy in the fluid rest-frame are shown from the top to bottom panels.}
	 \label{fig:rdrag}
\end{figure*}
Next, we solve a test problem suitable to examine the numerical implementation for the interaction process between radiation and fluid. The units of $c=a_{\rm rad}=1$ are employed in this subsection. In this test, the one-zone homogeneous medium initially moving toward the $x$-direction with $v^x=0.9$ is considered. The rest-mass density and pressure are initially set to be $\rho=1$ and $p=0.1$. $\Gamma_{\rm th}$ is set to be $5/3$. The initial radiation field is set to be uniform and isotropic with the energy density of $1.0$ in the laboratory frame. As is the case for Sec.~\ref{sec:test:therm}, a single cell is set for the grid. The periodic boundary condition is applied to the $\pm x$ boundaries. Three setups of the absorption and scattering opacity with $(\kappa_{\rm abs}, \kappa_{\rm sct})=(0.0,1.0)$, $(1.0,0.0)$, and $(0.1,0.9)$ are considered. For all the cases, we set $N_{\rm trg}=1.2\times 10^5$. The computation with 3 different time steps ($\Delta t=$1, 0.5, and 1/128) is performed. As a reference, the numerical solutions of Ref.~\cite{2020ApJ...901...96A} obtained by $\Delta t=0.01$ are also shown.

For this setup, the $x$-component of the radiation flux in the fluid rest-frame is initially negative. Hence, the fluid will be decelerated through the fluid-radiation interaction (i.e., radiation dragging), while the radiation flux in the fluid rest-frame approaches 0. Finally, the system will relax to a stationary state at the time at which the  radiation flux in the fluid rest-frame vanishes. 

Figure~\ref{fig:rdrag} show the results of the radiation-dragging test problems. For the pure scattering case, i.e., $(\kappa_{\rm abs}, \kappa_{\rm sct})=(1.0,0.0)$; see the left panel in Fig.~\ref{fig:rdrag}, we find that $e_{\rm rad}\neq T_{\rm gas}^4$, and this indicates that the thermal equilibrium is not achieved since there is no emission or absorption process which changes the photon number. On the other hand, for the pure absorption case, i.e., $(\kappa_{\rm abs}, \kappa_{\rm sct})=(0.0,1.0)$; see the middle panel in Fig.~\ref{fig:rdrag}, $e_{\rm rad}= T_{\rm gas}^4$ is achieved after relaxing to a stationary state, and this indicates that the thermal equilibrium is achieved. This is also the case for $(\kappa_{\rm abs}, \kappa_{\rm sct})=(0.1,0.9)$, although much longer time is needed until reaching the stationary state due to a small value of the absorption coefficient. 

For all the setups, we confirm that our code reproduces the results of Ref.~\cite{2020ApJ...901...96A}. More quantitatively, defining the L2 deviation of the result with $\Delta t=1/128$ with respect to that of Ref.~\cite{2020ApJ...901...96A} by
\begin{align}
 	\sqrt{\frac{\int\left|e_{\rm rad}(t)-e_{\rm rad,ref}(t)\right|^2 dt}{\int e_{\rm rad,ref}(t)^2 dt}}\label{eq:l2error2},
\end{align}
we find that our results agree with the results of Ref.~\cite{2020ApJ...901...96A} by $\lesssim 1\%$. Here, $e_{\rm rad}(t)$ and $e_{\rm rad,ref}(t)$ denote radiation energy density obtained by our $\Delta t=1/128$ run and in Ref.~\cite{2020ApJ...901...96A}, respectively. 

As is also seen in Sec.~\ref{sec:test:therm}, a long time scale is required for the computation with larger values of $\Delta t$ until the system with a finite value of absorption opacity relaxes to the thermal equilibrium state. Nevertheless, approximately the same asymptotic values are achieved for all the physical variables regardless of $\Delta t$ after the system is relaxed. 

\subsubsection{Radiation hydrodynamics shock-tube problem}
\begin{table*}
\begin{center}
\begin{tabular}{c|cccc|cccc}
	&&&$x<0$&&&&$x\ge0$&\\
	Model &	~~~$\rho$~~~&	$u^x$&	$p_{\rm gas}$&	$e_{\rm rad, com}$&	~~~$\rho$~~~&	$u^x$&	$p_{\rm gas}$&	$e_{\rm rad, com}$\\\hline
	non-relativistic &
1&	0.015&	$3\times10^{-5}$&	$10^{-8}$&	2.4&	$6.25\times10^{-3}$&	1.61&	$2.5\times10^{-7}$\\
	non-relativistic, radiation dominant	&
1&	0.69&	$6\times10^{-3}$&	$0.18$&	3.65&	$0.189$&	$3.59\times10^{-2}$&	$1.3$\\
	relativistic shock	&
1&	10&	$60$&	$2$&	8&	$1.25$&	$2.34\times10^{3}$&	$1.13\times10^{3}$\\\hline
\end{tabular}
\caption{Initial values for the rest-mass density, the $x$-component of four-velocity, fluid pressure, and the fluid rest-frame radiation energy employed for 1D radiation hydrodynamics  shock-tube problems.}
\label{tb:sht_param}
\end{center}
\end{table*}

\begin{figure*}
 	 \includegraphics[width=.329\linewidth]{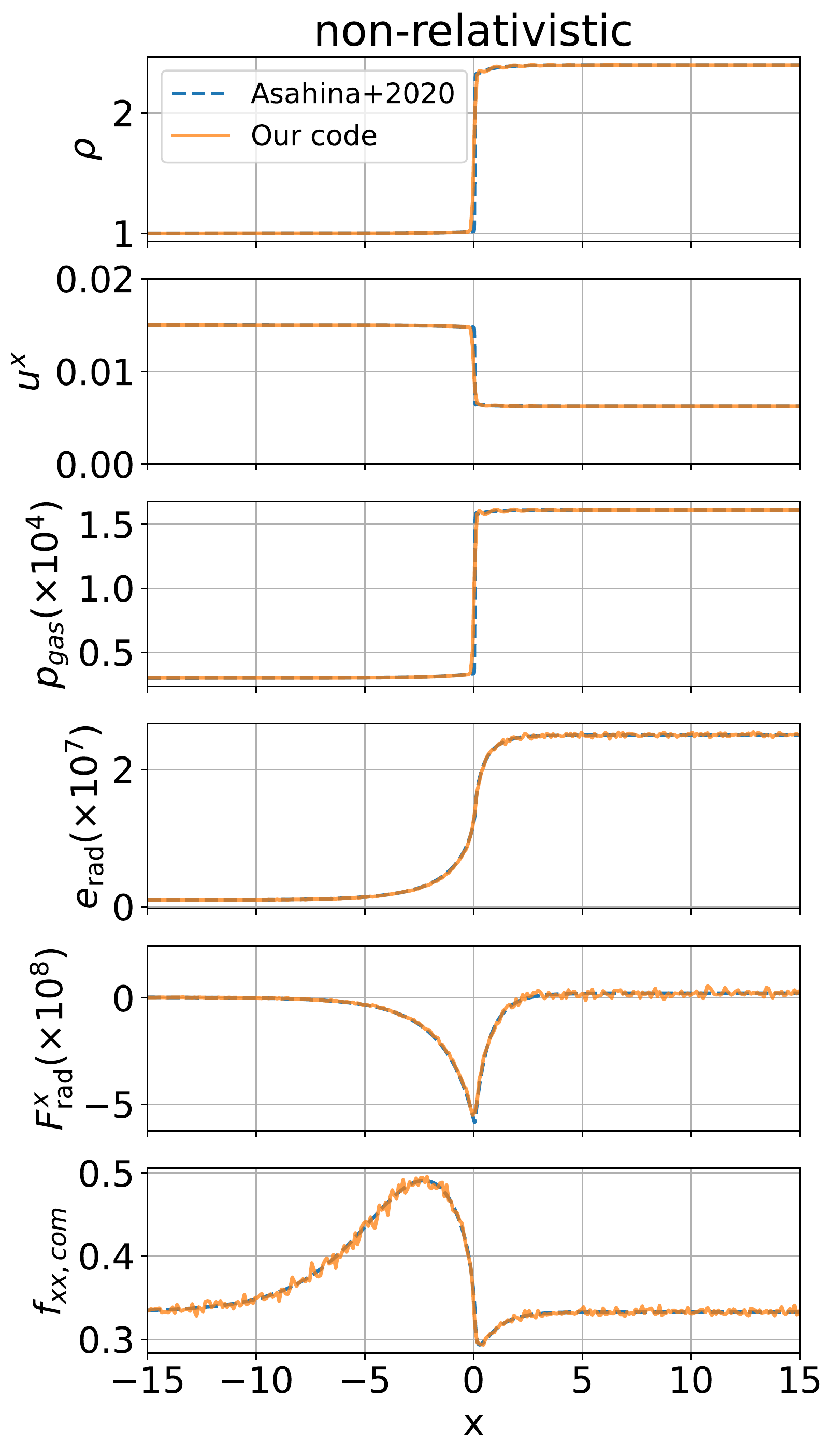}
 	 \includegraphics[width=.329\linewidth]{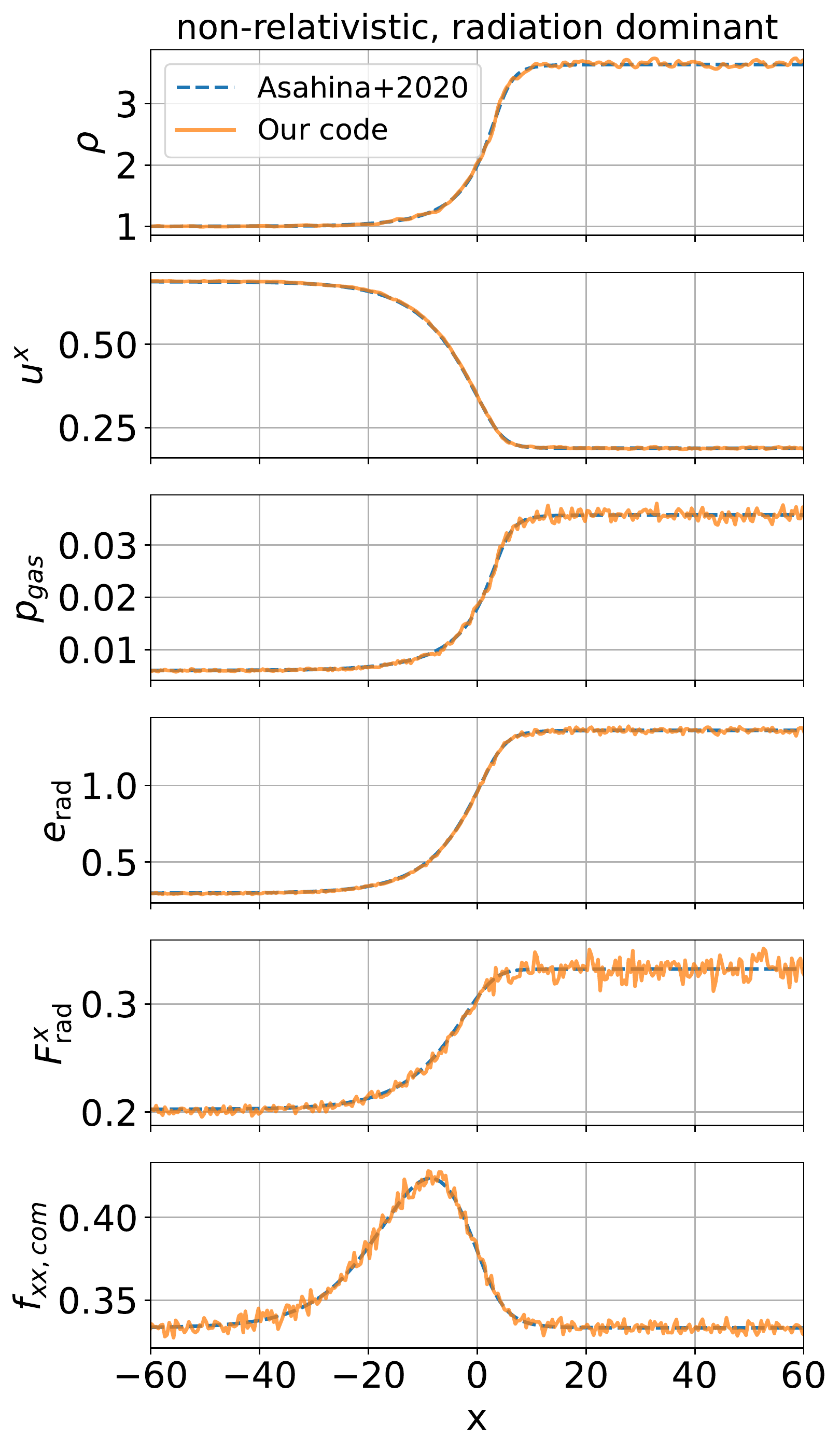}
 	 \includegraphics[width=.329\linewidth]{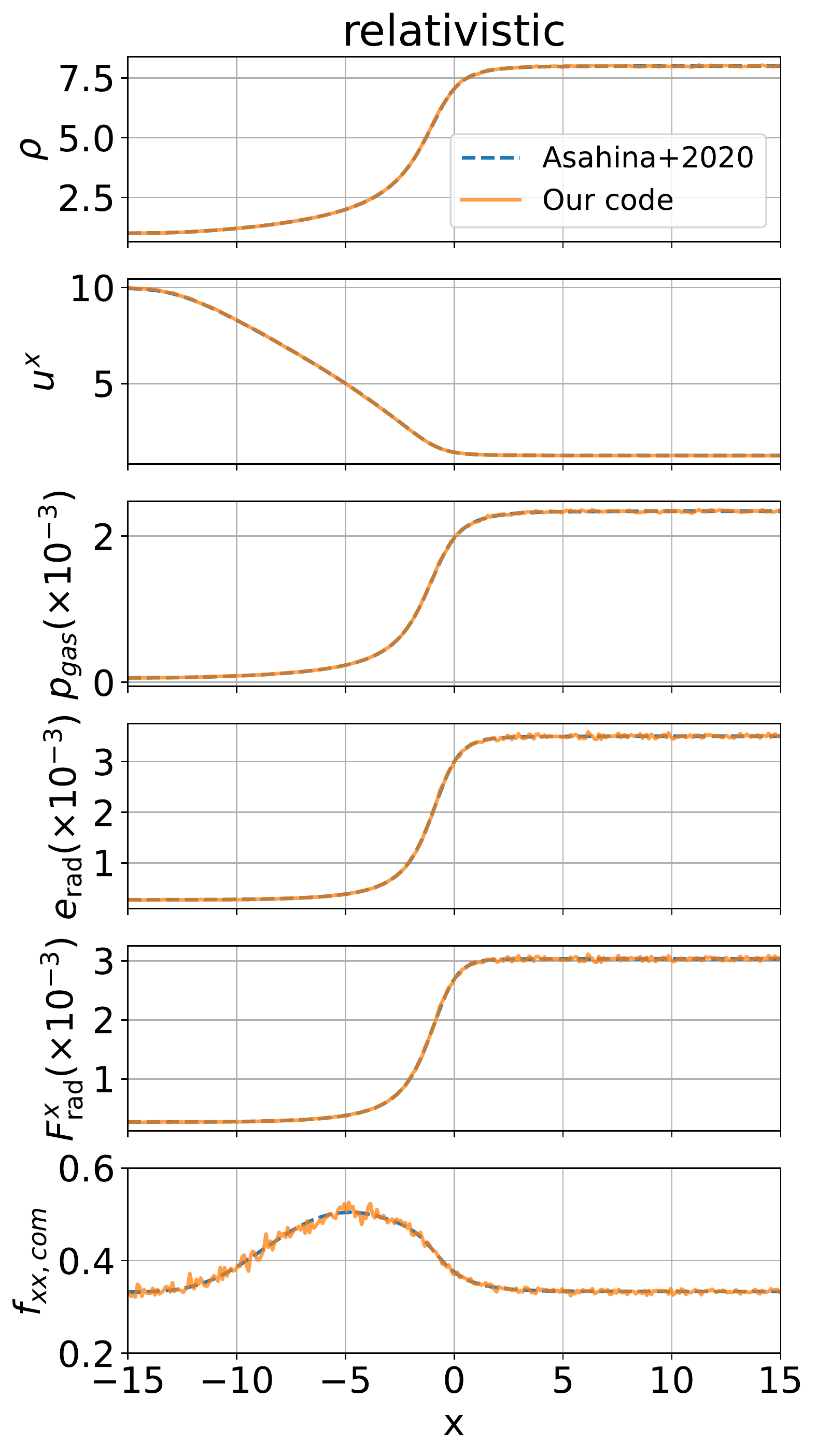}
 	 \caption{The rest-mass density, $x$-component of four-velocity, fluid pressure, laboratory-frame radiation-energy density, laboratory-frame radiation-energy flux, and $xx$-component of the Eddington tensor at $t=5000$ for the radiation hydrodynamics shock-tube problems. The left, middle, and right panel show the results for the non-relativistic, non-relativistic and radiation-pressure dominant, and relativistic shock-tube problems. The solid curves are our results and the dashed curves are those by Ref.~\cite{2020ApJ...901...96A}.}
	 \label{fig:sht-asahina}
\end{figure*}

In this subsection, we numerically solve the 1D shock-tube problems proposed in  Ref.~\cite{2008PhRvD..78b4023F} and performed in Refs.~\cite{2016ApJ...818..162O,2020ApJ...901...96A}. For this test, the same form of emissivity and equation of state as those in Sec.~\ref{sec:test:therm} are employed. 

Initially, the system is composed of 2 distinct homogeneous regions contacting at $x=0$ in the computational region of $[-L,L]$ with $L>0$ (see below). The radiation field is initially set to be isotropic. In this test, three setups of shock-tube problems are considered; one corresponds to a non-relativistic shock-tube problem, one corresponds to a non-relativistic and radiation-pressure dominant shock-tube problem, and one corresponds to a relativistic shock-tube problem. For the former two problems, $\Gamma_{\rm th}$ is set to be 5/3, while $\Gamma_{\rm th}=2$ is employed for the relativistic shock-tube problem. $\kappa_{\rm abs}$ for a non-relativistic, non-relativistic and radiation-pressure dominant, and relativistic shock-tube problems is set to be $0.4$, $0.08$, and $0.3$, respectively. The initial values for the rest-mass density, the $x$-component of four-velocity, fluid  pressure, and the fluid rest-frame radiation energy are summarized in Table~\ref{tb:sht_param}. Here, we note that the unit is chosen so that $c=1$ and $e_{\rm rad, com}=T_{\rm gas}^4$ are satisfied at the initial time.  For the non-relativistic, radiation-pressure dominant, and relativistic shock-tube problems, the simulations are performed for the region of $x=[-20,20]$, $[-80,80]$, and $[-20,20]$, respectively, and the number of the grid point is set to be $256$ for all the problems. Matter and radiation fields on the 4 grids at both edges of the computational domain are always reset to be identical with the initial states to obtain the stationary solutions. For all the runs, we employ $N_{\rm trg}=1.2\times 10^4$.

Figure~\ref{fig:sht-asahina} shows the results of the shock-tube problems at $t=5000$ and the comparison with those obtained in Ref.~\cite{2020ApJ...901...96A}. Here, the $xx$-component of the Eddington tensor is defined by $f_{\rm xx,com}=T_{\rm rad, com}^{xx}/T_{\rm rad,com}^{tt}$ using the energy momentum tensor of the radiation field measured in the fluid rest-frame, $T_{\rm rad,com}^{\mu\nu}$. We note that the shock front of our solution does not remain completely stationary during the evolution due to the numerical diffusion. Hence, the location of the coordinate origin for the results from Ref.~\cite{2020ApJ...901...96A} is slightly shifted with $\lesssim1$ to match our results.

It is shown that our code reproduces the results of Ref.~\cite{2020ApJ...901...96A} well. To check the agreement quantitatively, we calculate the L2 deviation of our solution from those obtained in Ref.~\cite{2020ApJ...901...96A} using the same definition as in Eq.~\eqref{eq:l2error1} but with $e_{\rm rad}(x)\rightarrow \rho(x)$ and $e_{\rm rad, analytic}(x)\rightarrow \rho_{\rm ref}(x)$ where $\rho(x)$ and $\rho_{\rm ref}(x)$ denote the rest-mass density obtained by our code and in Ref.~\cite{2020ApJ...901...96A}, respectively. The range of the integrals are taken to be $-3L/4\leq x \leq 3L/4$. We find that the L2 errors are $< 2\%$ for all the shock-tube problems. The errors are dominated by a small fluctuation of the profile induced by the Monte-Carlo shot noise.

\subsection{Optically thick shock}\label{sec:test:sh4}
In our code, the interaction between fluid and radiation is described as the feed-back of the emission, absorption, and scattering of the packets. On the other hand, in the optically thick limit, the fluid-radiation system can be treated as a single hydrodynamics system, and the effect of fluid-radiation interaction can be taken into account by considering the radiation pressure in the equation of state. In this test, we demonstrate that our code can capture the effect of fluid-radiation interaction properly even in an optically thick regime by showing that the result obtained for a very optically thick system agrees with the result obtained by a physically-equivalent pure hydrodynamics simulation in which the radiation pressure is included in the equation of state.

\begin{figure*}
 	 \includegraphics[width=.49\linewidth]{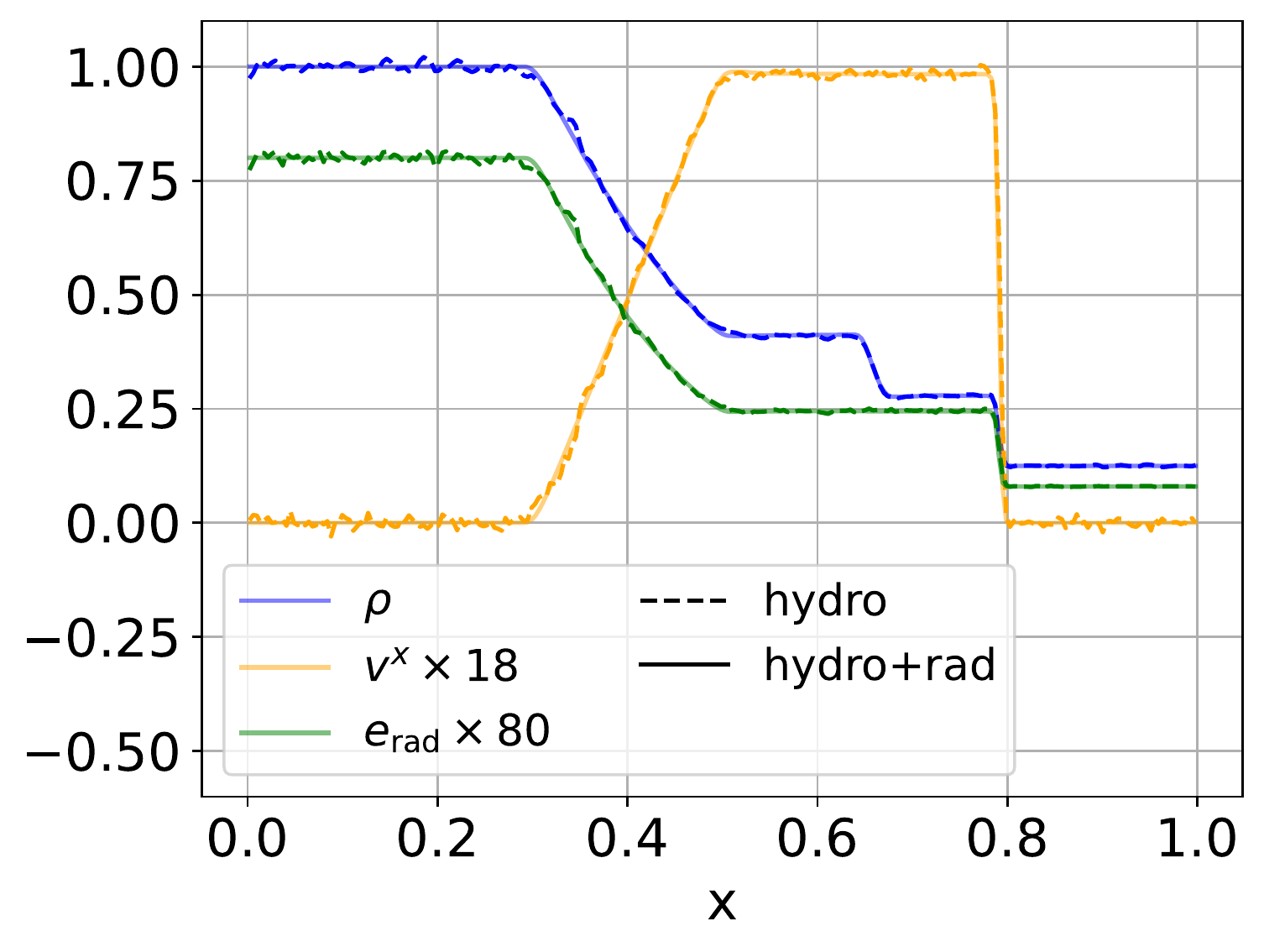}
 	 \includegraphics[width=.49\linewidth]{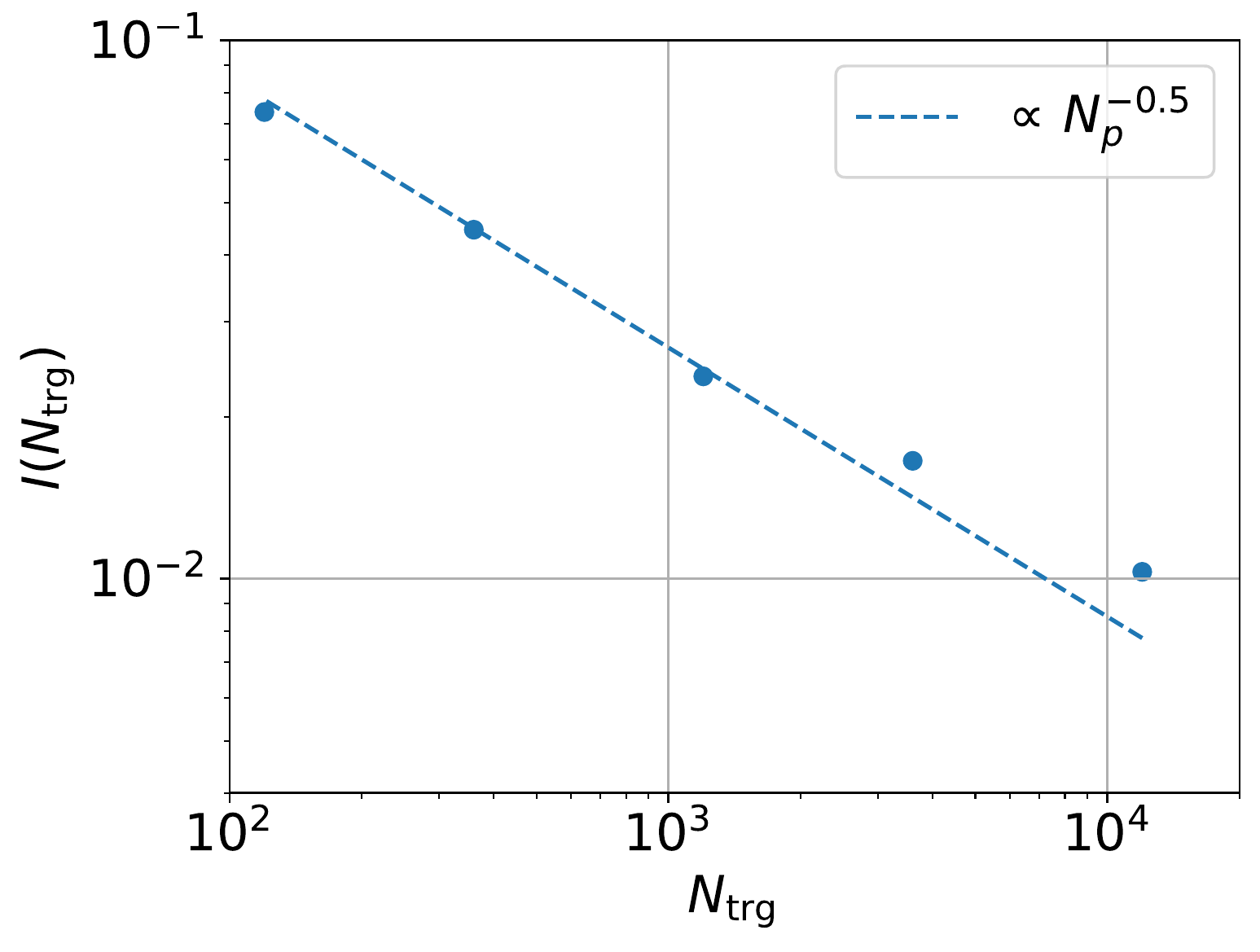}
 	 \caption{(Left panel) Rest-mass density, velocity, and radiation energy density at $t=3$ for the optically thick Sod's problem. The solid (``hydro+rad'') and dashed curves (``hydro'') show the results for the simulations in which the radiation field is solved by the Monte-Carlo scheme and by a pure hydrodynamics simulation, respectively. (Right panel) L2 deviation between the numerical solutions for the rest-mass density profile with finite values of 
 	  $N_{\rm trg}=120$, $360$, $1200$, $3600$, $12000$, and that at the $N_{\rm trg}\rightarrow \infty$ limit.}
	 \label{fig:sh_thick}
\end{figure*}

For a pure hydrodynamics simulation describing the fluid-radiation system in the optical thick limit, we employ the radiation-pressure dominant equation of state given by $P_{\rm gas}=\frac{1}{3}e_{\rm gas}$ where $e_{\rm gas}=a_{\rm rad} T_{\rm gas}^4$. (In this test, the continuity equation is solved although it is not necessary to determine the dynamics.) For the simulation in which the radiation-field sector is solved with the Monte-Carlo scheme, the equation of state of $P_{\rm gas}=\frac{1}{3}e_{\rm gas}$ with $e_{\rm gas}=f_{\rm eos} a_{\rm rad} T_{\rm gas}^4$ is employed to ensure that the same result is obtained as the pure hydrodynamics simulation in the optically thick limit. Indeed, by this choice of the equation of state, the total pressure will be the same as that in the pure hydrodynamics simulation for the same temperature after the thermal equilibrium state is reached. The ratio of fluid internal energy to radiation energy becomes $f_{\rm eos}/(1+f_{\rm eos})$ in the thermal equilibrium state. For simplicity, we here employ the units of $c=a_{\rm rad}=1$.

We consider a region of $0\leq x\leq 1$ with 256 grid points for the computation. Initially, the system is composed of two distinct homogeneous regions contacting at $x=0.5$. The fluid in these two regions is  initially at rest, and the rest-mass density and total internal energy density for $x<0.5$  and for $x\geq0.5$ are given by $(\rho,e_{\rm rad})=(1,0.01)$ and $(0.125,0.001)$, respectively. The reflective boundary condition is applied for both edges of the domain. $f_{\rm eos}$ is set to be 0.1. The initial radiation field is set so that the fluid and radiation fields are initially in the local thermal equilibrium. The absorption and scattering opacity is set to be $10^5$ and $0$, respectively. By these setups, the optical depth of each hydrodynamics cell is initially $\approx 390$ and $\approx49$ for $x\geq0.5$ and $x<0.5$, respectively.

The left panel of Fig.~\ref{fig:sh_thick} compares the rest-mass density at $t=3$ for a pure hydrodynamics simulation and the simulation in which the radiation field is solved by the Monte-Carlo scheme with $N_{\rm trg}=1.2\times 10^4$. This figure shows that the simulation in which the radiation field is solved by the Monte-Carlo scheme reproduces the result of the pure hydrodynamics simulation in the optically thick limit. This indicates that our code can appropriately solve fluid-radiation interaction for an optically thick system. 

Since the system is closed, the total energy and momentum perpendicular to the $x$-direction should be conserved. We find that the total energy is conserved within the relative error of $10^{-11}$, which is approximately to the level of the machine precision. The total azimuthal angular momentum, which is approximately 0 at $t=0$, is also conserved within the error of $10^{-14}\,E_{\rm tot}$ with $E_{\rm tot}$ being the total energy (remind that the computation is practically performed in an axisymmetric domain in which the $x$-axis is identified with the $z$-axis). 

To check the convergence property of the solution with respect to the Monte-Carlo packet number, we perform the simulations for the same optically thick shock-tube problem with various values of $N_{\rm trg}$ and calculate the L2 deviation between the solution for the rest-mass density profile with a finite value of $N_{\rm trg}$, $\rho_{N_{\rm trg}}(x)$, and that for $N_{\rm trg}\rightarrow \infty$ limit, $\rho_{\rm c}(x)$, defined by
\begin{align}
I(N_{\rm trg})=\sqrt{\frac{\int_0^1 \left[\rho_{N_{\rm trg}}(x)-\rho_{\rm c}(x)\right]^2 dx}{\int_0^1 \rho_{\rm c}(x)^2 dx}}.\label{eq:L2_sh4}
\end{align}
For $\rho_{\rm c}(x)$, we employ the rest-mass density profile of the pure hydrodynamics simulation, as we expect that it represents the solution at the $N_{\rm trg}\rightarrow \infty$ limit.

The right panel of Fig.~\ref{fig:sh_thick} shows the results of $I(N_{\rm trg})$ for  simulations with $N_{\rm trg}=120$, $360$, $1200$, $3600$, and $12000$. The sequence of $I(N_{\rm trg})$ is approximately proportional to $N_{\rm trg}^{-1/2}$ as is expected~\citep{2009ApJS..184..387D,2015ApJ...807...31R}. The small deviation from $\propto N_{\rm trg}^{-1/2}$ seen for $N_{\rm trg}\ge 3600$ may reflect the fact that the opacity is large but still finite while the solution of the pure hydrodynamics simulation is expected to be the limit of infinite opacity.

By the setting of $\tau_{\rm therm}=3$, the prescription introduced in Sec.~\ref{sec:thmreg} is not switched on in the computation even for the optically-thick shock-tube problem presented above due to the small value of the optical depth in the cells. Hence, to examine the prescription of Sec.~\ref{sec:thmreg}, we compute the same optically-thick shock-tube problem but with $\kappa_{\rm abs}=10^6$. The computation is performed for the cases of $\tau_{\rm therm}=1$ and $\tau_{\rm therm}\gg1$ (i.e., without applying the prescription), and of $N_{\rm trg}=1.2\times 10^3$ for both cases. We find that the L2 deviations computed by Eq.~\eqref{eq:L2_sh4} for these setups are $\approx0.015$ and $\approx0.027$, while the numbers of packets per cell are $\approx2200$ and $\approx1300$, respectively (note that the increase in the number of packets for the computation with $\tau_{\rm therm}=1$ is due to the additional packet creation/removal process of the prescription). This implies that the improvement of the L2 deviation for the computation with $\tau_{\rm therm}=1$ is equivalent to that achieved by increasing the number of packets by a factor of $\approx 4$ (see Fig.~\ref{fig:sh_thick}), while the actual increase in the number of the packets is only by a factor of $\approx 2$. 

\subsection{Convergence order with respect to the time resolution}\label{sec:test:conv}
\subsubsection{the one-zone problems}
\begin{figure*}
 	 \includegraphics[width=0.49\linewidth]{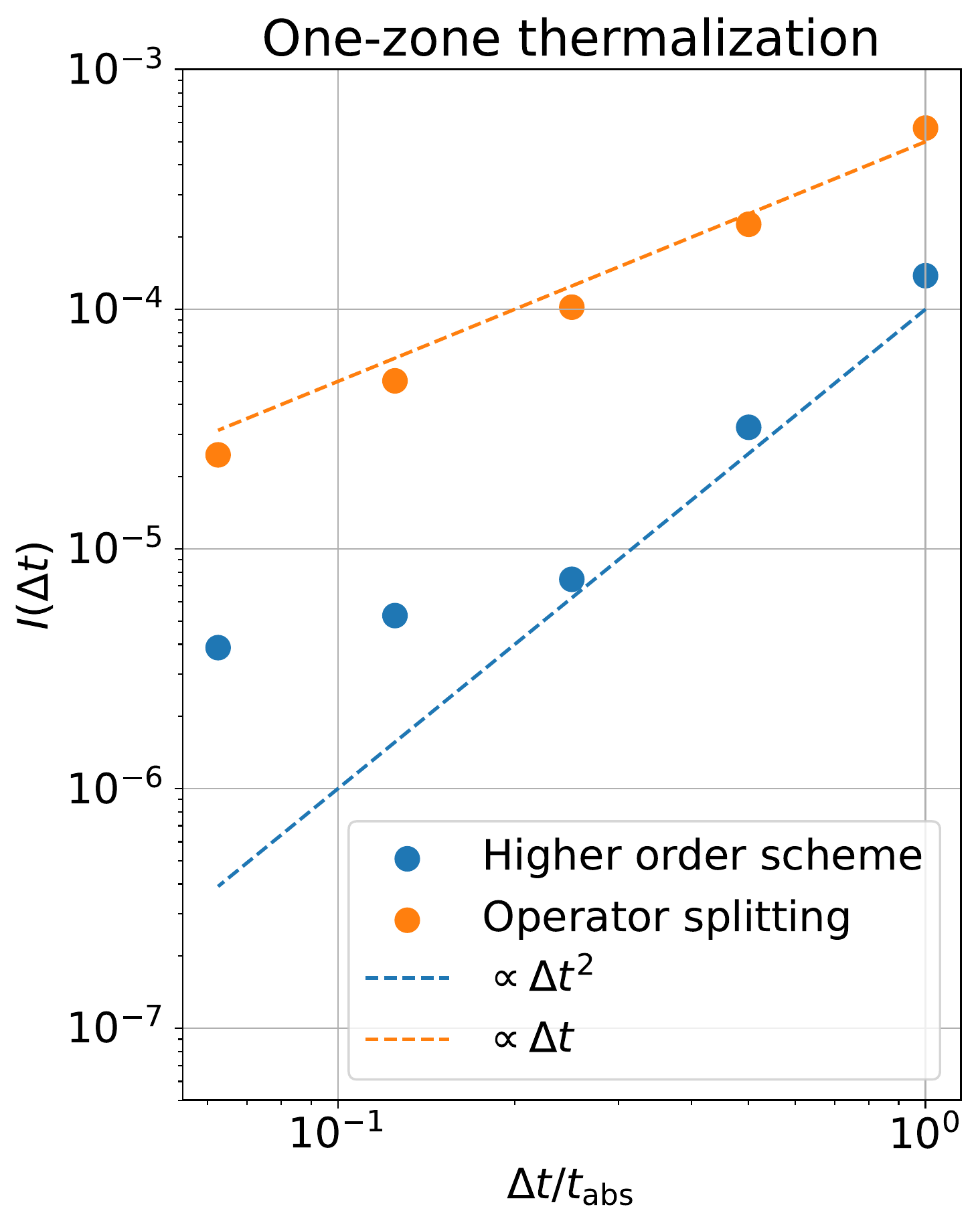}
 	 \includegraphics[width=0.49\linewidth]{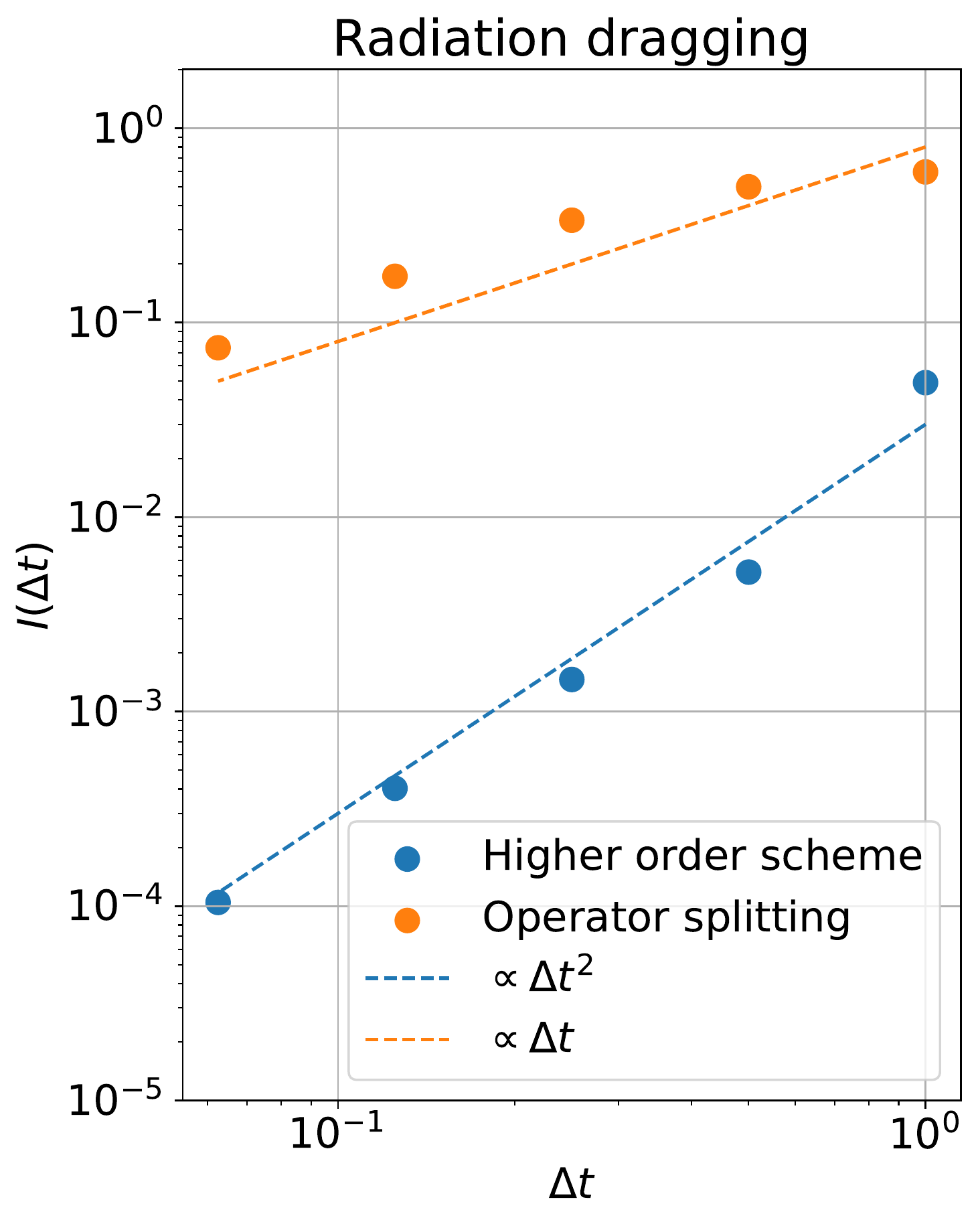}
 	 \caption{L2 errors for various time resolution runs for the fluid-pressure dominant case of the one-zone thermalization test (left panel) and pure scattering case of radiation-dragging test (right panel) performed in Sec.~\ref{sec:test:therm} and Sec.~\ref{sec:test:rdrag}, respectively.}
	 \label{fig:conv}
\end{figure*}

To demonstrate that a higher-order accuracy in time is indeed achieved in our code, we show the convergence property of the solutions for the fluid pressure dominant case of the one-zone thermalization test and for the pure scattering case of radiation dragging test performed in Sec.~\ref{sec:test:therm} and Sec.~\ref{sec:test:rdrag}, respectively.

To evaluate the convergence property of the numerical solutions, we calculate the L2 errors of the solutions with various values of $\Delta t$, which is defined by
\begin{align}
	I(\Delta t)=\sqrt{\frac{
	\sum_{k=0}^{n}\left[T^{\Delta t}_{\rm gas}\left(k\Delta t_0\right)-T^{\rm 0}_{\rm gas}\left(k\Delta t_0\right)\right]^2
	}
	{
	\sum_{k=0}^{n}T^{\rm  0}_{\rm gas}\left(k\Delta t_0\right)^2
	}}.
\end{align}
Here, $T^{\Delta t}_{\rm gas}(t)$ denotes the numerical solutions of fluid temperature for the test problems obtained by a finite value of $\Delta t$ and, $T^{0}_{\rm gas}(t)$ denotes the numerical solutions of the test problems at the $\Delta t\rightarrow0$ limit. $\Delta t_0$ denotes the largest value of $\Delta t$, and we consider the numerical solution for $t=[0,n\Delta t_0]$. For the one-zone thermalization test and radiation dragging test, we set $\Delta t=t_{\rm abs}$ and 1, respectively, with $n=5$ for both cases. Since it is practically difficult to obtain the exact solutions of the test problems at the $\Delta t\rightarrow0$ limit, we approximate $T^{0}_{\rm gas}(t)$ with a numerical solution obtained with $\Delta t= \Delta t_0/128$. We set $N_{\rm trg}$ to be  $1.2\times 10^7$ and $1.2\times 10^6$ for  the one-zone thermalization test and radiation dragging test, respectively. The large values of $N_{\rm trg}$ are used to suppress the Monte-Carlo shot noise and to focus only on the error induced by the finite time resolution.
 
Figure~\ref{fig:conv} shows the results of $I\left(\Delta t\right)$ for $\Delta t/\Delta t_0=1$, $1/2$, $1/4$, $1/8$, $1/16$, and $1/32$. This shows that the sequence of $I\left(\Delta t\right)$ is approximately proportional to $\Delta t^{2}$ for both test problems if $\Delta t/\Delta t_0\gtrsim 0.2$. If the numerical error of the solution for $T_{\rm gas}$ due to the finite time resolution is approximately proportional to $\Delta t^{p}$ with the convergence order, $p$, for a sufficiently small value of $\Delta t$, we expect that $I(\Delta t)$ is also proportional to $\Delta t^{p}$.  Hence, the behavior of $I\left(\Delta t\right)\propto \Delta t^{2}$ suggests that the second-order accuracy is indeed achieved with respect to the time resolution.

As the reference, we also show the results in which an operator splitting scheme is employed. For this case, we clearly find that $I\left(\Delta t\right)$ is approximately proportional to $\Delta t$ for both test problems, suggesting that the operator splitting scheme results in the first-order accuracy in time for problems in which the  fluid-radiation interaction is important. The error of the result obtained by the operator splitting scheme is always larger than that by the higher-order scheme. We should emphasize that this is the case even if the value of $\Delta t$ is comparable with the physical time scale of the system. This implies that our higher-order scheme may be useful not only to improve the convergence property of the numerical solution but also to suppress the numerical error for problems in which a large value of $\Delta t$ is inevitable.

We note that the convergence order decreases to $\lesssim0.5$ for $\Delta t/\Delta t_0\lesssim 0.2$ for the one-zone thermalization test. This reflects the fact that the error is dominated by that due to the shot noise of the Monte-Carlo packets. Hence, considering the computational costs, we should keep in mind that the advantages of employing the higher-order time integration scheme may diminish for the case that the Monte-Carlo shot noise dominates the numerical error. We should also give a caution that the second-order accuracy is not always achieved for a large value of $\Delta t$, particularly, for the region in which the implicit Monte-Carlo scheme plays an important role. The reason is simply due to the fact that the assumption that the numerical error of the solution due to the finite time resolution is proportional to $\Delta t^{p}$, which is based on the Taylor expansion of the solution with respect to $\Delta t$, is no longer valid if the value of $\Delta t$ is larger than a typical evolution time scale of the system. Indeed, the implicit Monte-Carlo scheme did not play an important role in the fluid-pressure dominant case of the one-zone thermalization test and pure scattering case of radiation-dragging test. 

\subsubsection{1D clump problem}
\begin{figure*}
 	 \includegraphics[width=0.49\linewidth]{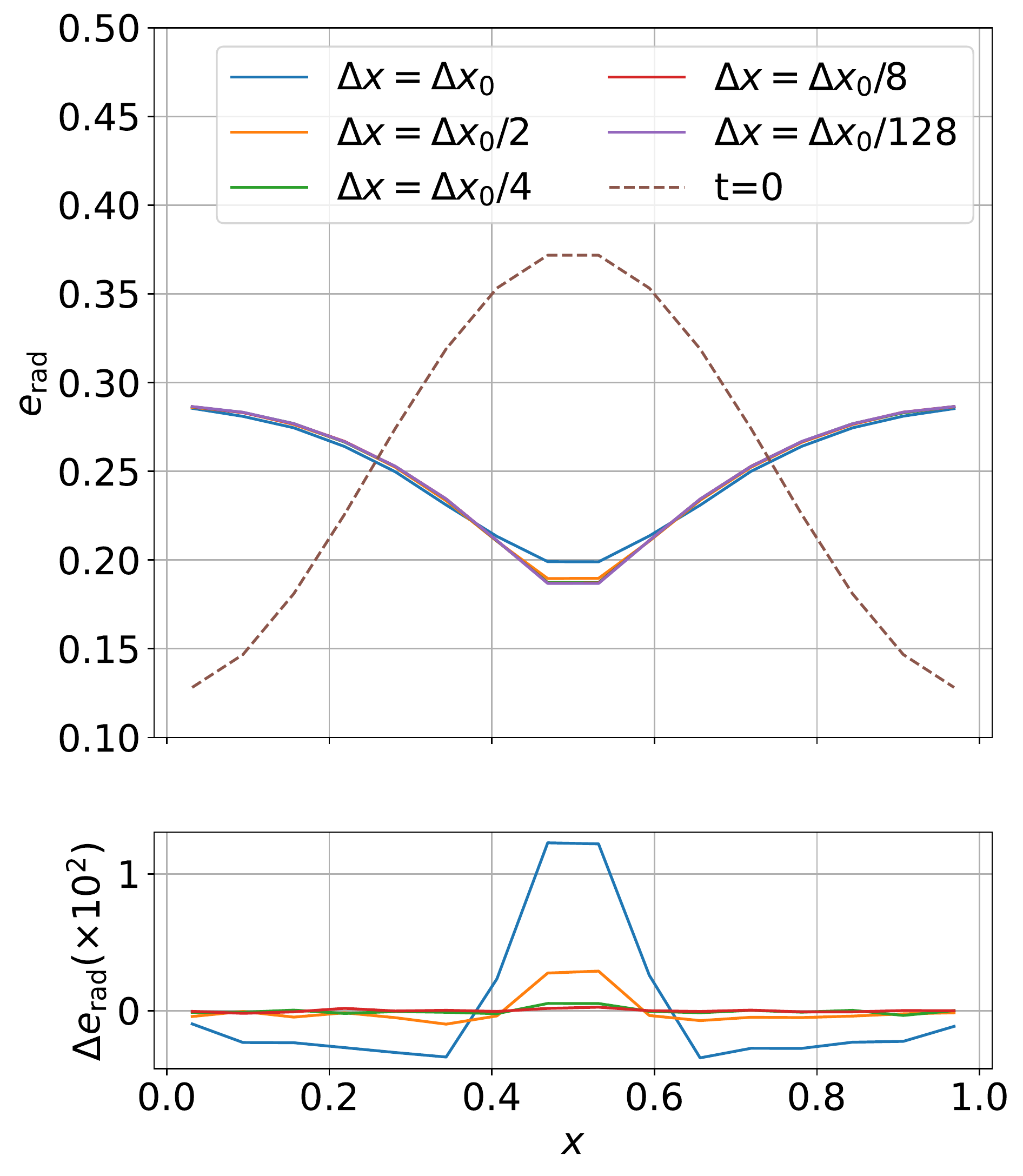}
 	 \includegraphics[width=0.49\linewidth]{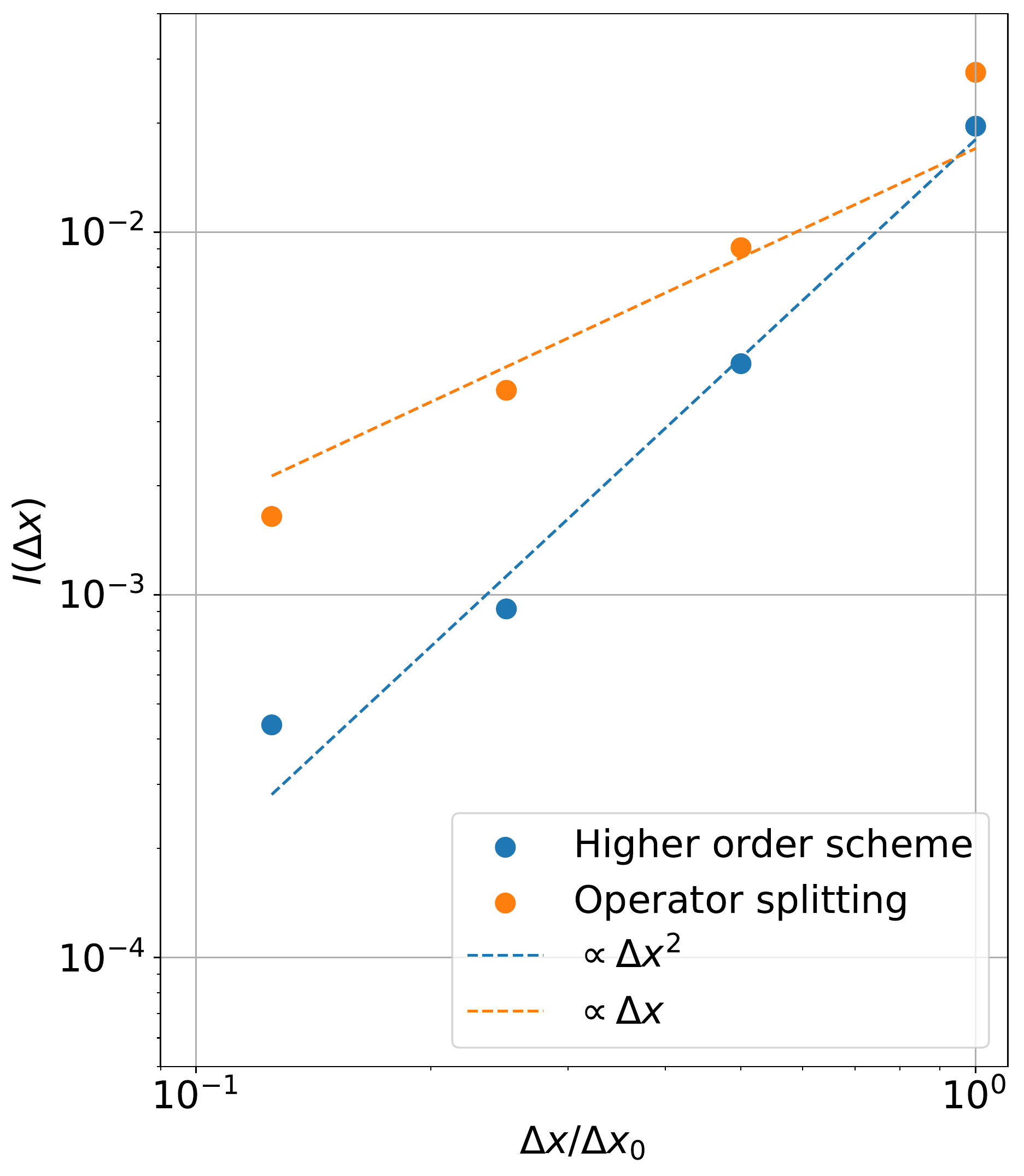}
 	 \caption{(Left panel) Laboratory-frame radiation energy density at $t=1.56$ for various grid setups. The deviation of the result for each grid setup from that for the finest grid setup, i.e., the one with 2048 grid points, is shown in the left-bottom panel. (Right panel) L2 errors of the laboratory-frame radiation energy density for the 1D density clump problem with various grid spacing.}
	 \label{fig:conv2}
\end{figure*}

In addition to the second-order accuracy in time, the radiative processes are considered up to the second-order accuracy in space in our code. To demonstrate that our code has indeed the second-order accuracy in both time and space, we solve the following 1D density clump problem with various time and spatial grid resolution. Here we consider a region of $0\leq x\leq 1$ with the periodic boundary condition, and set the initial rest-mass density given by $\rho(x)=1-0.5\,{\rm cos}\left(2\pi x\right)$. The system is initially at rest ($v^x=0$), and the total specific internal energy is set to be uniformly $1$. The same units and equation of state as in Sec.~\ref{sec:test:sh4} but with $f_{\rm eos}=3$ are employed. Radiation is initially set to be in thermal equilibrium. Absorption and scattering opacity is set to be $10$ and $0$, respectively. In this problem, shock waves, in the presence of which the hydrodynamics solver will be first-order accurate, do not appear and the dynamical time scale can be easily resolved so as to justify the Taylor expansion of the solution. Hence, this problem is suitable for examining our higher-order time integration scheme.

The number of the grids is set to be 16, 32, 64, 128, and 2048, and the grid spacing $\Delta x$ for them is $\Delta x/\Delta x_0=1,1/2,1/4,1/8,$ and $1/128$, respectively, with $\Delta x_0$ being the grid spacing of the computation with 16 grid points. We choose $\Delta t=0.5\Delta x$, and hence, the computation with a finer grid setup is evolved with a smaller value of $\Delta t$. $N_{\rm trg}$ is set to be $1.2\times 10^6$ except for the computation with $\Delta x/\Delta x_0=1/128$, for which $N_{\rm trg}=1.2\times 10^5$ is employed. In order to compare the profile of fluid and radiation among different grid resolutions, all the profiles of the physical variables are averaged in the spatial bins which agree with the grid structure of the run with $\Delta x=\Delta x_0$.

The left panel of Figure~\ref{fig:conv2} shows the laboratory-frame radiation energy density at $t=1.56$ for various grid setups. Radiation is partially trapped by the fluid and such a component follows the fluid motion, while gradually becoming more homogeneous due to diffusion. A clump is initially present around $x=0.5$ and expands with time. Subsequently, a dimple is formed at the location of the initial peak. The time of the snapshot shown in the left panel of Fig.~\ref{fig:conv2} corresponds to the time at which a dimple is formed for the first time. The deviation of the result for each grid setup from those for the finest grid setup (i.e., the one with 2048 grid points) is shown in the left bottom part of Fig.~\ref{fig:conv2}. We find that the results converge to the finest-grid one as the grid resolution is improved. 

To discuss the convergence property of the solution, quantitatively, we calculate the L2 deviation between the solution with a finite value of $\Delta x$, $e_{{\rm rad}, \Delta x}(x)$, and that at the limit of $\Delta x\rightarrow \infty$, $e_{{\rm rad}, c}(x)$, defined by
\begin{align}
I(\Delta x)=\sqrt{\frac{\int_0^1 \left[e_{{\rm rad}, \Delta x}(x)-e_{{\rm rad}, c}(x)\right]^2 dx}{\int_0^1 e_{{\rm rad}, c}(x)^2 dx}}.
\end{align}
Practically, we use the profile at $t=1.56$ and approximate $e_{{\rm rad}, c}(x)$ by the result with $\Delta x/\Delta x_0=1/128$. The right panel of Fig.~\ref{fig:conv2} shows the results of $I\left(\Delta x\right)$ for $\Delta x/\Delta x_0=$1, 1/2, 1/4, and 1/8. As the reference, the results obtained by employing the operator splitting scheme are also shown. The sequence of $I\left(\Delta x\right)$ obtained by the higher-order scheme is approximately proportional to $\Delta x^{2}$. Exceptionally, 
for $\Delta x/\Delta x_0=1/8$, we obtain a larger value of $I\left(\Delta x\right)$ than the trend of $\propto \Delta x^{2}$. In this case, the error is likely to be dominated by the Monte-Carlo shot noise. 

The results employing the operator splitting scheme are approximately proportional to $\Delta x$. Since the implementation of the code other than the time integration part is identical in between the higher-order scheme and operator splitting scheme, this result suggests that the L2 deviation employing the operator splitting scheme is determined primarily by the time integration error.

\begin{figure*}
 	 \includegraphics[width=.329\linewidth]{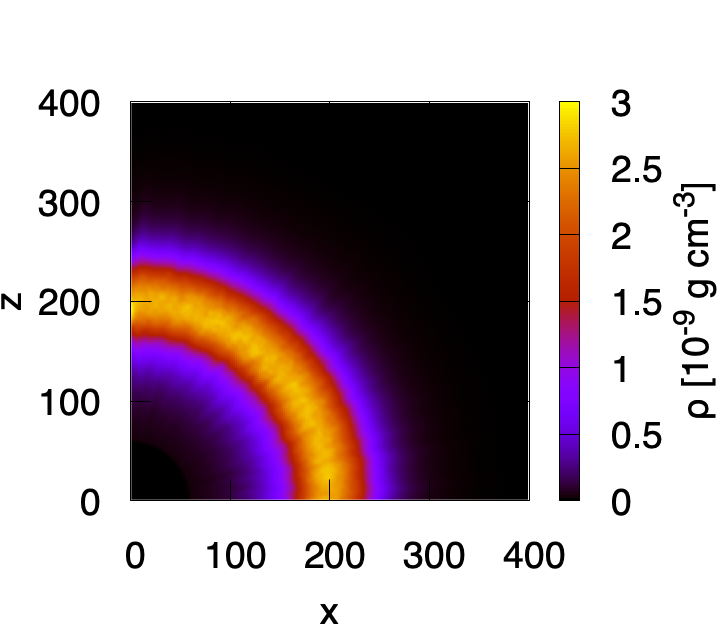}
 	 \includegraphics[width=.329\linewidth]{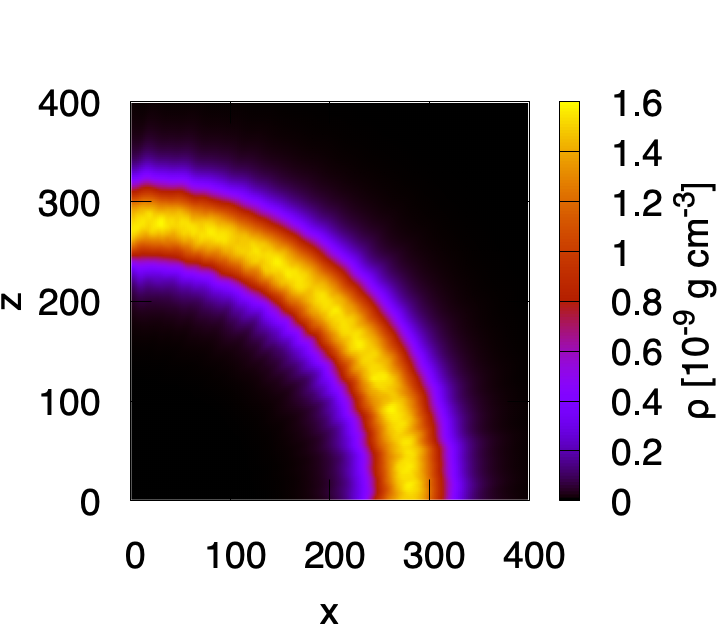}
 	 \includegraphics[width=.329\linewidth]{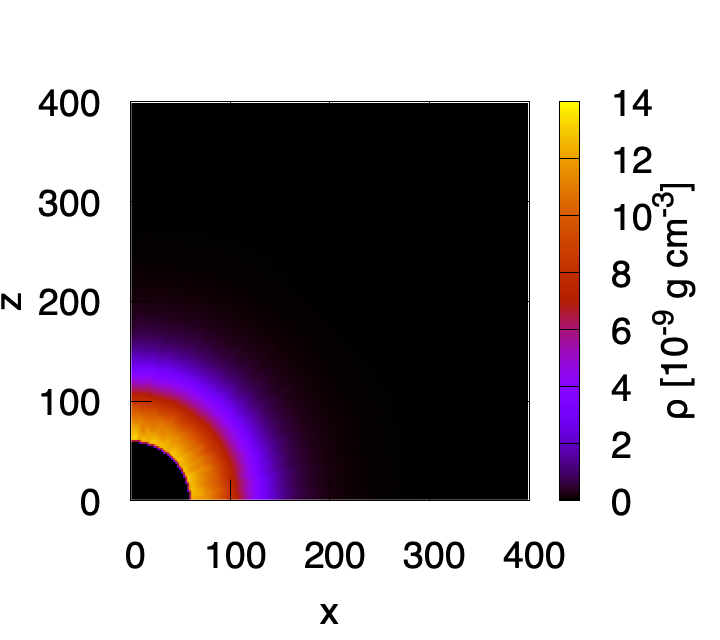}
 	 \caption{Rest-mass density profile of the Eddington limit tests at $t=t_{\rm ff}$. The left, middle, and right panels show the results for $f_{\rm inj}=1$, 1.5, and 0.5, respectively.}
	 \label{fig:edd}
\end{figure*}

\begin{figure}
 	 \includegraphics[width=\linewidth]{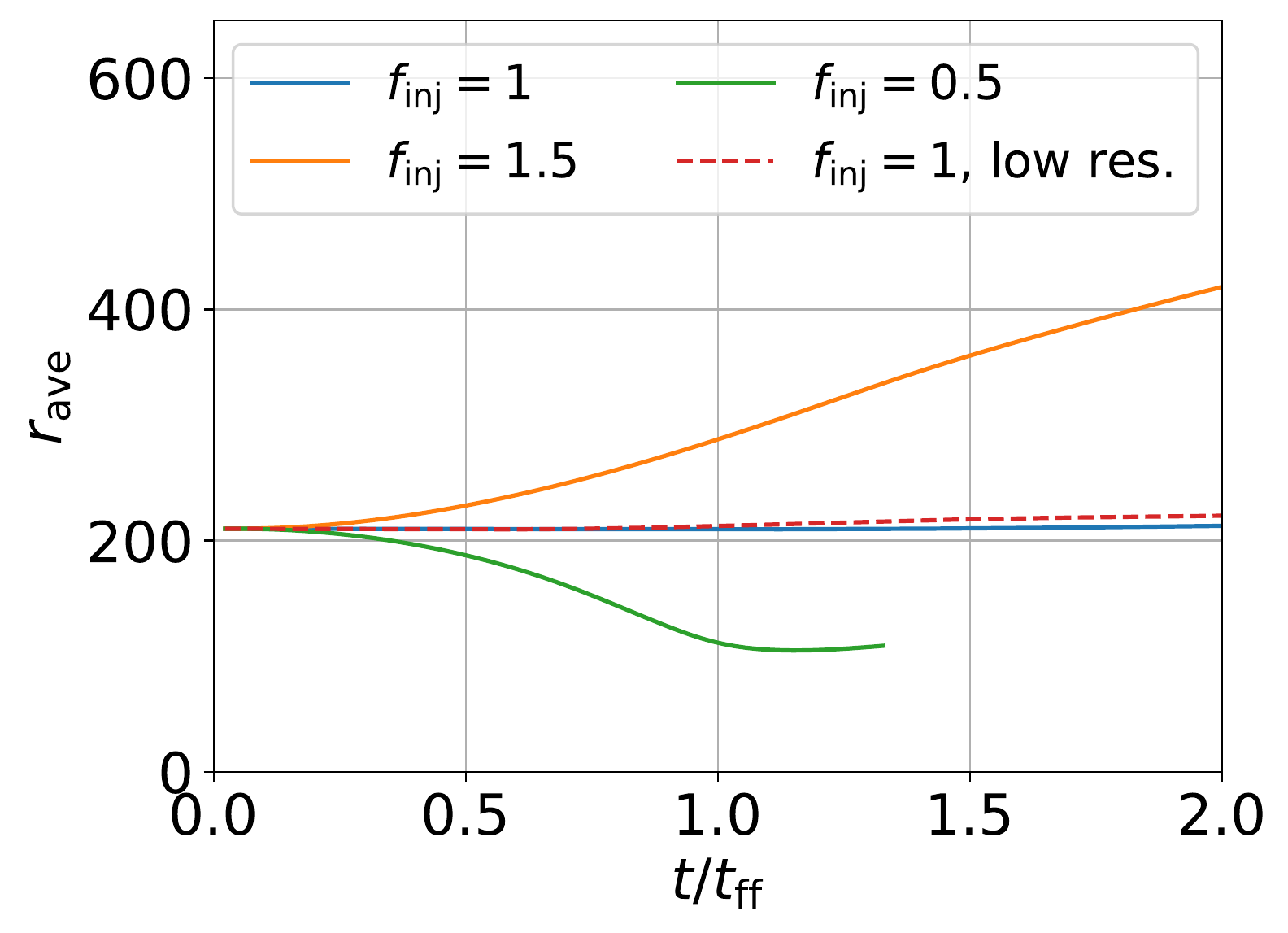}
 	 \caption{Averaged radius of the shell determined by the rest-mass density average. The results for $f_{\rm inj}=$ 1, 1.5, and 0.5 are shown. For $f_{\rm inj}=1$, the result calculated with a low grid resolution ($100\times100$ grids) and small packet number for injected radiation ($1.2\times 10^4$ packets per a time step) is also shown (``low res.''). The result for $f_{\rm inj}=0.5$ is truncated at the time at which the $90\%$ of the shell mass fall into the region of $r<60\,r_{\rm g}$ because the matter which reaches the region is removed from the simulated region and the value of the averaged radius is not meaningful any longer.}
	 \label{fig:edd_rave}
\end{figure}
\subsection{Eddington limit test}
As the final test, we examine whether the Eddington limit is captured by our code for axisymmetric problems in the presence of gravitation. In this test,  we prepare a  non-rotating black-hole metric with the black-hole mass of $M_{\rm BH}=1\,M_\odot$ in the isotropic coordinates as the fixed background. The simulated region is set to be $0\leq x\leq400\,r_{\rm g}$ and $0\leq z\leq400\,r_{\rm g}$, where $r_{\rm g}=G M_{\rm BH}/c^2$,  imposing the equatorial symmetry in addition to axisymmetry. A uniform grid with the number of 200 is set for both $x$ and $z$ directions. We initially set a spherical shell at $200\leq r/r_{\rm g}\leq 210$ where $r$ is the radial coordinate of the isotropic coordinates. The rest-mass density and absorption opacity are set to be $10^{-8}\,{\rm g\,cm^{-3}}$ and $1.5\times10^{-8}\,{\rm cm^2\,g^{-1}}$, respectively, and the scattering opacity is set to be 0. The same emissivity and equation of state as in Sec.~\ref{sec:test:therm} but with $\Gamma_{\rm th}=1.01$ are employed so that the thermal expansion of the shell should be negligible. We inject photons from $r=50\,r_{\rm g}$ with the luminosity measured in the spatial infinity to be $L=f_{\rm inj}L_{\rm inj}=4\pi f_{\rm inj} GM_{\rm BH}/\kappa_{\rm abs}c$. $2.4\times 10^4$ packets are injected for each time step, while $N_{\rm trg}$ is set to be $120$ because the emission from the shell is not important. We do not solve the evolution of the hydrodynamics sector until $t=200\,r_{\rm g}/c$ in order to wait for the spreading of the injected radiation over the simulation region. The matter that falls into the region of $r<60\,r_{\rm g}$ is removed so as not to disturb the photon injection. $r_{\rm abs}=0.1$ is employed to reduce the Monte-Carlo shot noise during the absorption process.

In the absence of injected radiation, the shell should fall into the central black hole in the time scale of the free-fall, $t_{\rm ff}=\pi/2\sqrt{G M_{\rm BH}/r^3}\approx4500\,r_{\rm g}/c$. In the presence of the injected radiation, the momentum is transported into the shell by the absorption of radiation and the infall motion is suppressed. For $f_{\rm inj}\approx 1$, we expect that the center of the shell approximately keeps the initial radius by the balance between the gravitational and radiation forces. For $f_{\rm inj}\gtrsim 1$, the radiation force overcome the gravitational force, and hence, the shell should expand outwards.

To examine that our code can reproduce these physical results, we perform the simulations with $f_{\rm edd}=0.5$, $1$, and $1.5$. Figure~\ref{fig:edd} shows the rest-mass density profile at $t=t_{\rm ff}$ and Fig.~\ref{fig:edd_rave} the averaged radius of the shell determined by the rest-mass density average. For $f_{\rm edd}=1$, we find that the shell approximately keeps the initial location at least for the free-fall time scale. We confirm that this property does not change significantly even if we employ lower grid resolutions or smaller packet numbers for injected radiation (see the results with the label of ``low res.'' in Fig.~\ref{fig:edd_rave}). For $f_{\rm edd}=0.5$, the shell falls into the central region, although the infalling speed is slower than that in the absence of the injected radiation. For $f_{\rm edd}=1.5$, by contrast, the shell is pushed outward by the radiation pressure and moves to larger radii. All these results are consistent with the physical expectation.

\section{Summary}\label{sec:sum}
In this paper, we presented our new Monte-Carlo-based relativistic radiation hydrodynamics code. Our code is developed based on the previous works~\citep{2009ApJS..184..387D,2012ApJ...755..111A,2015ApJS..217....9R,2015ApJ...807...31R,2018MNRAS.475.4186F}, but in addition, we proposed the following new ingredients for the Monte-Carlo scheme in this paper: 
\begin{itemize}

\item We proposed and implemented a new procedure to achieve the second-order accuracy for the time integration in the limit of a large packet number even in the presence of significant matter--radiation interaction. In this higher-order time integration scheme the energy-momentum conservation is guaranteed to the precision of the geodesic integrator. 

\item The spatial dependence of radiative processes, such as the packet propagation, emission, absorption, and scattering, is taken into account up to the second-order accuracy.

\item We proposed a new method to determine the Fleck parameter, $\alpha_{\rm eff}$, which is a key variable to control the effective opacity introduced in the implicit Monte-Carlo scheme. The new implementation is a generalization of the originally and previously employed one of choosing the parameter. It has an advantage that the updated radiation and fluid energy do not overshoot those of the local thermal equilibrium state even if they are initially far from the equilibrium. 

\item We proposed and implemented a prescription to skip the evolution of the packet deep inside the cell in which the thermal equilibrium is likely to be achieved. By this prescription the Monte-Carlo shot noise of the energy-momentum transport between the cells is reduced for the fixed number of packets due to an effective increase of the packet density in the region near the cell interface.

\end{itemize}

We validated our code by reproducing the solutions of various test-problems following the previous studies; one-zone thermalization, dynamical diffusion, radiation dragging, radiation mediated shock-tube, shock-tube in the optically thick limit, and Eddington limit problems. By comparing our numerical results with the exact solutions and/or the numerical solutions obtained in Ref.~\cite{2020ApJ...901...96A}, we confirmed that our code can reproduce the solutions for a number of the test problems with reasonable accuracy. We also demonstrated  that the energy-momentum conservation is achieved to the level of the machine precision for the case of flat spacetime.

We demonstrated that the second-order accuracy is indeed achieved with our higher-order time integration scheme for one-zone and 1D problems. We also reconfirmed that the computation based on the operator splitting scheme results in the first order accuracy in time. We found that the error of the result obtained by the operator splitting scheme is always larger than that by the higher-order scheme even for the case that the time step is comparable with the physical time scale of the system. This suggests the merit of employing our higher-order scheme to suppress the numerical error accompanied with a large time-step size. On the other hand, the second-order accuracy in time is not always achieved in the presence of a large Monte-Carlo shot noise or for a large time step, i.e., for the case that the implicit Monte-Carlo scheme plays an essential role. Hence, whether the higher-order time integration scheme should be employed or not depends on the problem and computational resources.

There are several tasks remaining for the development of our code. One is the implementation of realistic microphysics, such as the equations of state, emissivity, and opacity. Although the implementation of them is rather straightforward, we should examine whether we can stably solve the system even in the presence of their complicated dependence on temperature, rest-mass density, and electron fraction. Implementation of neutrino-antineutrino pair annihilation process in dynamical spacetime is also a target of our future development. The implementation of the discrete diffusion technique~\citep{2015ApJS..217....9R,2015ApJ...807...31R,2018PhRvD..98f3007F} will be a great help to reduce the computational costs in a highly scattering regime. Both hydrodynamics and radiation solvers of our code is parallelized under OpenMP, but efficient parallelization with MPI computing has to be achieved to solve problems with larger grid/packets numbers. 

\acknowledgments KK thanks Tomohisa Kawashima and Katsuaki Asano for the valuable discussions. KK thanks Yuta Asahina and Ken Ohsuga for providing us their numerical solution of test problems. This work was supported by Grant-in-Aid for Scientific Research (JP21K13912 and JP20H00158) of JSPS/MEXT.

\appendix

\section{Higher-order time integration}\label{app:hoti}
In this section, we show that the higher-order time integration scheme introduced in this work is indeed accurate to the second-order in time in the limit of a large packet number.

As in Sec.~\ref{secIIIE}, we consider the time evolution of matter and radiation fields given by ${\bf u}(t)$ and ${\bf y}(t)$, respectively. Generally, the time derivatives of ${\bf u}(t)$ and ${\bf y}(t)$ are functions of ${\bf u}(t)$ and ${\bf y}(t)$, and thus, the  basic equations are written schematically as 
\begin{align}
\frac{d{\bf y}(t)}{dt}=F\left[{\bf y}(t),{\bf u}(t)\right],\\
\frac{d{\bf u}(t)}{dt}=G\left[{\bf y}(t),{\bf u}(t)\right].
\end{align}
The evolution of the radiation field is given by solving the propagation, creation, and annihilation of the consisting packets over the fixed matter field. Hence, the new radiation field ${\bf y}_1$ obtained by solving the evolution of the packets under the fixed matter field of ${\bf u}_0$ can be considered as the solution of the radiation field in which the time evolution of the matter field is neglected. Formally, this can be described by
\begin{align}
{\bf y}_1={\bf y}_0+\int_0^{\Delta t} ds\,F\left[{\bf y}|_{{\bf u}={\bf u}_0}(t+s),{\bf u}_0\right].\label{eq:y10}
\end{align}
Here, ${\bf y}|_{{\bf u}={\bf u}_0}(t+s)$ denotes the solution of the radiation field of which the state is ${\bf y}_0$ at $t=0$ and the matter field is virtually fixed as ${\bf u}_0$ in the time evolution. The evolution of ${\bf y}_0\rightarrow {\bf y}_1$ is accurate in time to the order of the time integration scheme employed for solving the packet propagation. In the following, we assume that the time integration for the packet evolution is accurate at least up to the second order. Then, we can expand Eq.~\eqref{eq:y10} with respect to $\Delta t$ as 
\begin{align}
{\bf y}_1={\bf y}_0+F\left[{\bf y}_0,{\bf u}_0\right] \Delta t+\frac{1}{2}\frac{\delta F}{\delta {\bf y}} F\left[{\bf y}_0,{\bf u}_0\right]\Delta t^2+{\cal O}(\Delta t ^3).\label{eq:y1}
\end{align}

The radiation feedback calculated during the time evolution of ${\bf y}_0\rightarrow{\bf y}_1$ is used to evolve the matter field. We can schematically write this as
\begin{align}
{\bf u}_1={\bf u}_0+\int_0^{\Delta t} ds\,G\left[{\bf y}|_{{\bf u}={\bf u}_0}(t+s),{\bf u}_0\right]. 
\end{align}
Again, the radiation feedback obtained here can also be considered to be as accurate in time as ${\bf y}_1$ supposing that the matter field is virtually fixed as ${\bf u}_0$. Then, we have
\begin{align}
{\bf u}_1={\bf u}_0+G\left[{\bf y}_0,{\bf u}_0\right] \Delta t+\frac{1}{2}\frac{\delta G}{\delta {\bf y}}F\left[{\bf y}_0,{\bf u}_0\right] \Delta t^2+{\cal O}(\Delta t ^3).\label{eq:u1}
\end{align}

In the next sub-step, the matter and radiation fields are evolved in the same way as in the first step but using ${\bf u}_1$ as the fixed matter field. Let ${\bf y}_2$ and ${\bf u}_2$ be radiation and matter fields obtained by this second sub-step, respectively. Substituting ${\bf y}_1$ and ${\bf u}_1$ of Eqs.~\eqref{eq:y1} and~\eqref{eq:u1} into the evolution equations, ${\bf y}_2$ and ${\bf u}_2$ can be expressed as
\begin{align}
{\bf y}_2&={\bf y}_0+F\left[{\bf y}_0,{\bf u}_1\right] \Delta t+\frac{1}{2}\frac{\delta F}{\delta {\bf y}} F\left[{\bf y}_0,{\bf u}_1\right] \Delta t^2+{\cal O}(\Delta t ^3)\nonumber\\
&={\bf y}_0+F\left[{\bf y}_0,{\bf u}_0\right] \Delta t+\frac{\delta F}{\delta {\bf u}}G\left[{\bf y}_0,{\bf u}_0\right] \Delta t^2\nonumber\\
&+\frac{1}{2}\frac{\delta F}{\delta {\bf y}}F\left[{\bf y}_0,{\bf u}_0\right] \Delta t^2+{\cal O}(\Delta t ^3),
\end{align}
\begin{align}
{\bf u}_2&={\bf u}_0+G\left[{\bf y}_0,{\bf u}_1\right] \Delta t+\frac{1}{2}\frac{\delta G}{\delta {\bf y}}F\left[{\bf y}_0,{\bf u}_1\right] \Delta t^2+{\cal O}(\Delta t ^3)\nonumber\\
&={\bf u}_0+G\left[{\bf y}_0,{\bf u}_0\right] \Delta t+\frac{\delta G}{\delta {\bf u}} G\left[{\bf y}_0,{\bf u}_0\right]\Delta t^2\nonumber\\
&+\frac{1}{2}\frac{\delta G}{\delta {\bf y}}F\left[{\bf y}_0,{\bf u}_0\right] \Delta t^2+{\cal O}(\Delta t ^3),
\end{align}
respectively.

In the last sub-step, the evolution of the matter and radiation fields is calculated in the same ways as in the first and second steps but using $\frac{1}{2}{\bf u}_*=\frac{1}{2}{\bf u}_0+\frac{1}{4}{\bf u}_1+\frac{1}{4}{\bf u}_2$ as the fixed matter field. Following the same procedure as for calculating ${\bf y}_2$ and ${\bf u}_2$, ${\bf y}_3$ and ${\bf u}_3$ are expressed as
\begin{align}
{\bf y}_3&={\bf y}_0+F\left[{\bf y}_0,{\bf u}_*\right] \Delta t+\frac{1}{2}\frac{\delta F}{\delta {\bf y}} F\left[{\bf y}_0,{\bf u}_0\right]\Delta t^2+{\cal O}(\Delta t ^3)\nonumber\\
&={\bf y}_0+F\left[{\bf y}_0,{\bf u}_0\right] \Delta t+\frac{1}{2}\frac{\delta F}{\delta {\bf u}}G\left[{\bf y}_0,{\bf u}_0\right] \Delta t^2\nonumber\\
&+\frac{1}{2}\frac{\delta F}{\delta {\bf y}}F\left[{\bf y}_0,{\bf u}_0\right] \Delta t^2+{\cal O}(\Delta t ^3),
\end{align}
\begin{align}
{\bf u}_3&={\bf u}_0+G\left[{\bf y}_0,{\bf u}_*\right] \Delta t+\frac{1}{2}\frac{\delta G}{\delta {\bf y}}F\left[{\bf y}_0,{\bf u}_*\right] \Delta t^2+{\cal O}(\Delta t ^3)\nonumber\\
&={\bf u}_0+G\left[{\bf y}_0,{\bf u}_0\right] \Delta t+\frac{1}{2}\frac{\delta G}{\delta {\bf u}} G\left[{\bf y}_0,{\bf u}_0\right]\Delta t^2\nonumber\\
&+\frac{1}{2}\frac{\delta G}{\delta {\bf y}}F\left[{\bf y}_0,{\bf u}_0\right] \Delta t^2+{\cal O}(\Delta t ^3),
\end{align}
respectively.

Finally, the radiation and matter fields for the next step, ${\bf y}_{\rm new}$ and ${\bf u}_{\rm new}$, are determined by the following relations:
\begin{align}
{\bf y}_{\rm new}=\frac{1}{6}{\bf y}_1+\frac{1}{6}{\bf y}_2+\frac{2}{3}{\bf y}_3,
\end{align}
\begin{align}
{\bf u}_{\rm new}=\frac{1}{6}{\bf u}_1+\frac{1}{6}{\bf u}_2+\frac{2}{3}{\bf u}_3,
\end{align}
respectively.
Then we have
\begin{align}
{\bf y}_{\rm new}&={\bf y}_0+F\left[{\bf y}_0,{\bf u}_0\right] \Delta t+\frac{1}{2}\frac{\delta F}{\delta {\bf u}}G\left[{\bf y}_0,{\bf u}_0\right] \Delta t^2\nonumber\\
&+\frac{1}{2}\frac{\delta F}{\delta {\bf y}}F\left[{\bf y}_0,{\bf u}_0\right] \Delta t^2+{\cal O}(\Delta t ^3),
\end{align}
\begin{align}
{\bf u}_{\rm new}&={\bf u}_0+G\left[{\bf y}_0,{\bf u}_0\right] \Delta t+\frac{1}{2}\frac{\delta G}{\delta {\bf u}}G\left[{\bf y}_0,{\bf u}_0\right] \Delta t^2\nonumber\\
&+\frac{1}{2}\frac{\delta G}{\delta {\bf y}}F\left[{\bf y}_0,{\bf u}_0\right] \Delta t^2+{\cal O}(\Delta t ^3).
\end{align}
We can easily show that ${\bf y}_{\rm new}$ and ${\bf u}_{\rm new}$ agree with the expansion of ${\bf y}(t+\Delta t)$ and ${\bf u}(t+\Delta t)$ to second order with $\Delta t$, respectively. Furthermore, we confirm that ${\bf u}_{\rm new}$ agrees with that obtained by the so-called SSPRK3 method~\citep{1998MaCom..67...73G}. Hence, this time integration scheme is accurate up to the third order in time for the hydrodynamics sector in the limit of the negligible radiation feedback.

\section{Thinning and joining of the radiation fields}\label{app:pmix}
In this section, we describe the thinning method which is used for the higher-order time integration scheme in our code. 
 
We explain our method starting from a simple example. Let ${\bf y}_1$ and ${\bf y}_2$ be the radiation fields, which are both ${\bf y}_0$ at the initial time $t$ but obtained by the time evolution of $\Delta t$ under two different matter fields. Let ${\cal U}_1$ (${\cal U}_2$) be a set of packets, which contains packets in ${\bf y}_0$ and those generated during the time evolution of ${\bf y}_0\rightarrow{\bf y}_1$ (${\bf y}_0\rightarrow{\bf y}_2$) between $t$ and $t+\Delta t$. Let $\Delta G^j_{1,k}$ ($\Delta G^j_{2 ,k}$) be the radiation four-force to the $j$-th cell which is induced by the time evolution of the $k$-th packet in ${\cal U}_1$ (${\cal U}_2$). The total radiation force to the $j$-th hydrodynamics cell is then described by $\Delta G^j_1=\sum_{k\in {\cal U}_1} \Delta G^j_{1,k}$ ($\Delta G^j_2=\sum_{k\in {\cal U}_2} \Delta G^j_{2,k}$). 

For the case that a new radiation field is constructed by ${\bf y}'=\lambda {\bf y}_1+(1-\lambda){\bf y}_2$, the packets in ${\bf y}_1$ and ${\bf y}_2$ are thinned out by fractions of $\lambda$ and $1-\lambda$, respectively, with respect to each total packet number. If we regard ${\bf y}_0\rightarrow{\bf y'}$ as the time evolution to the next time step, the radiation force $\{\Delta G'^j\}$ during the time evolution should be determined consistently with the packets included in ${\bf y'}$ to guarantee the energy-momentum conservation. For instance, the thinned out packets which are not picked up from ${\bf y}_1$ or ${\bf y}_2$ for constructing ${\bf y'}$ should not contribute to $\{\Delta G'^j\}$. In other words, for any contribution to $\{\Delta G'^j\}$, a packet which induces such radiation feedback should be included in ${\bf y'}$ unless absorbed during the time evolution. We note that the radiation feedback from the packets in ${\cal U}_1$ or ${\cal U}_2$ which are created and already absorbed during the time evolution should also be taken into account with the proportion of $\lambda$ and $1-\lambda$, respectively. 

For this purpose, we divide ${\cal U}_1$ (${\cal U}_2$) into subsets of ${\cal U}_{1A}$ and ${\cal U}_{1B}$ (${\cal U}_{2A }$ and ${\cal U}_{2B}$) for which the ratio of the packet number between ${\cal U}_{1A}$ and ${\cal U}_{1B}$ (${\cal U}_{2A }$ and ${\cal U}_{2B}$) is $\lambda:1-\lambda$. Then, the new radiation field ${\bf y}'$ and radiation four-force induced during ${\bf y}_0\rightarrow{\bf y'}$ are given by ${\bf y}'=({\bf y}_1\cap{\cal U}_{1A})\cup({\bf y}_2\cap{\cal U}_{2B})$, $\Delta G'^j=\sum_{k\in {\cal U}_{1A}}  \Delta G^j_{1,k}+\sum_{k\in {\cal U}_{2B}} \Delta G^j_{2,k}$. In this way, each contribution to $\Delta G'^j$ has a corresponding packet in ${\cal U}_{1A}$ or ${\cal U}_{2B}$ which induces such radiation force, and hence, the energy-momentum conservation of the system is naturally guaranteed. Practically, the division of the packet sets (${\cal U}_1={\cal U}_{1A}\bigoplus{\cal U}_{1B}$ and ${\cal U}_2={\cal U}_{2A}\bigoplus{\cal U}_{2B}$) can be done before evolving the radiation field, since the packets generated in the period from $t$ to $t+\Delta t$ can be predetermined at the beginning of the time evolution.

We note that the packets in the initial configuration ${\bf y}_0$ should be divided in the same way for ${\cal U}_1$ and ${\cal U}_2$; it is desirable to divide the packets as ${\cal U}_1={\cal U}_{1A}\bigoplus{\cal U}_{1B}$ and ${\cal U}_2={\cal U}_{2A}\bigoplus{\cal U}_{2B}$ so that ${\bf y}_0\cap{\cal U}_{1A}={\bf y}_0\cap{\cal U}_{2A}$ and ${\bf y}_0\cap{\cal U}_{1B}={\bf y}_0\cap{\cal U}_{2B}$. By this way, all the packets initially included in ${\bf y}_0$ will contribute to ${\bf y}'$ or $\Delta G'^j$, while a fraction of the packets experiences the evolution of ${\bf y}_0\rightarrow{\bf y}_1$ and the other fraction experiences ${\bf y}_0\rightarrow{\bf y}_2$. In particular, if the space time is stationary and there is no packet generation or annihilation, i.e., no interaction with the matter during the evolution, the resultant ${\bf y}'$ will be the same as that without applying the higher-order time integration scheme, because the time evolution of ${\bf y}_0\rightarrow{\bf y}_1$ and ${\bf y}_0\rightarrow{\bf y}_2$ is identical except for the probabilistically determined part.

The actual code is a little bit more complicated due to the presence of the Runge-Kutta sub-steps, but the procedure is essentially the same as explained above. First, at the beginning of the $i$-th Runge-Kutta sub-step, all the packets at the initial time $t$ and the packets which are created in the time evolution, ${\cal U}_i$, are divided into sets of ${\cal U}^0_i$, ${\cal U}^{(1)}_i$, ${\cal U}^{(2)}_i$, ${\cal U}^{(3)}_i$, ${\cal U}^{(4)}_i$, ${\cal U}^{(5)}_i$. Here, ${\cal U}^{(m)}_i$ ($m=0,1,2,3,4,5$) is determined so that $\#{\cal U}^{(0)}_i:\#{\cal U}^{(1)}_i:\#{\cal U}^{(2)}_i:\#{\cal U}^{(3)}_i:\#{\cal U}^{(4)}_i:\#{\cal U}^{(5)}_i=\frac{1}{6}:\frac{1}{3}:\frac{1}{12}:\frac{1}{6}:\frac{1}{12}:\frac{1}{6}=2:4:1:2:1:2$, where $\#$ denotes the number of the packets. For the case that the number of the packet is not a multiple of 12, the remainder is assigned to either of sets probabilistically following the packet number weight (however, as we discuss below, for the packets newly created in each time step, we require those number to be a multiple of 12 to have the correct emission rate). 

Then, the radiation field is evolved following the packet transport, creation,  and annihilation. The radiation field after the evolution (${\bf y}^{(m)}_i$) and the radiation four-force ($\Delta G^{(m)}_i$) induced during the evolution are computed for each $m$ in each Runge-Kutta sub-step. In the first Runge-Kutta sub-step, the total induced radiation four-force determined during the evolution is simply used to calculate the matter field at the next sub-step. In the second Runge-Kutta sub-step, a composite value of the radiation four-force is needed to calculate ${\bf u}_*=\frac{1}{2}{\bf u}_0+\frac{1}{4}{\bf u}_1+\frac{1}{4}{\bf u}_2$. This is calculated from $\Delta G^*=(\Delta G^{(2)}_1+\Delta G^{(3)}_1)+(\Delta G^{(4)}_2+\Delta G^{(5)}_2)$. After the third Runge-Kutta sub-step finished, the radiation field and radiation four-force obtained in the first, second, and third Runge-Kutta sub-steps are combined with the ratio of $\frac{1}{6}:\frac{1}{6}:\frac{2}{3}$, respectively, to obtain the radiation field of the next time step (${\bf y}_{\rm new}$) and radiation four-force ($\Delta G$). Specifically, they are obtained by ${\bf y}_{\rm new}=({\bf y}_1\cap{\cal U}^{(3)}_{1})\cup({\bf y}_2\cap{\cal U}^{(5)}_{2})\cup({\bf y}_3\cap{\cal U}^{(0)}_{3})\cup({\bf y}_3\cap{\cal U}^{(1)}_{3})\cup({\bf y}_3\cap{\cal U}^{(2)}_{3})\cup({\bf y}_3\cap{\cal U}^{(4)}_{3})$ and $\Delta G=\Delta G^{(3)}_1+\Delta G^{(5)}_2+(\Delta G^{(0)}_3+\Delta G^{(1)}_3+\Delta G^{(2)}_3+\Delta G^{(4)}_3)$. The radiation tensor and the energy of the packet that escapes from the system can be synthesized in the same way as the radiation reaction.

In this algorithm, the number of the packet created in each time step should be a multiple of 12 to ensure that expected emission rate is derived. For instance, if only one packet is created in the first Runge-Kutta sub-step and if it is assigned to the group other than ${\cal U}^{(3)}_1$ probabilistically, the contribution of the emission in this sub-step to the new radiation field (${\bf y}_{\rm new}$) and radiation four-force ($\Delta G$) cannot be taken into account. On the other hand, if the packet is assigned to ${\cal U}^{(3)}_1$, the effect of emission to ${\bf y}_{\rm new}$ and $\Delta G$ is overestimated in this sub-step (remind that ideally only $1/6$ of emission in this sub-step should contribute to the evolution). Hence, although the effect of emission is consistent with the desired value in probability average, the remainder at the packet division can cause an artificial fluctuation in emissivity. 

To avoid the artificial fluctuation in emissivity, in our implementation, the number of the created packet is adjusted to be a multiple of 12 to the utmost extent in the way explained in Sec.~\ref{sec:method:ems}. However, if the number of the created packet is small and it is numerically inefficient to adjust the number to be a multiple of 12, we give up applying the higher-order time integration scheme to the packet creation in the cell. Instead, we apply the following procedure at the time of the packet division:
\begin{enumerate}
	\item In the first Runge-Kutta sub-step, we assign all the created packets to ${\cal U}^{(0)}_1$.
	\item In the second Runge-Kutta sub-step, we assign all the created packets to ${\cal U}^{(4)}_2$ with the packet weight being enhanced by a factor of 2.
	\item In the third Runge-Kutta sub-step, we assign all the created packets to ${\cal U}^{(0)}_3$.
\end{enumerate}
In this way, the contribution of the created packet in each sub-step to the radiation field and radiation feedback will be limited to these update in the same sub-step without being lost, and thus, we can avoid the unphysical fluctuation in emissivity. However, we note that, by this prescription, the time integration accuracy will drop to the first order. 

\bibliographystyle{apsrev4-2}
\bibliography{ref}

\end{document}